\documentstyle[prc,aps,epsf,bbold]{revtex}
\newcommand {\beq}
            {\begin{equation}}
\newcommand {\beqa}
            {\begin{eqnarray}}
\newcommand {\eeq}
            {\end{equation}}
\newcommand {\eeqa}
            {\end{eqnarray}}
\begin {document}
\begin{title}
{
\hfill{\small {\bf MKPH-T-99-7.}}\\
{\bf Complete Sets of Polarization Observables in Electromagnetic Deuteron
Break-up}
\footnote{Dedicated to Professor Walter Gl\"ockle on the occasion of his
60th birthday}
\footnote{Supported by the Deutsche 
Forschungsgemeinschaft (SFB 443) and the National Science and 
Engineering Research Council of Canada}
}
\end{title}
\bigskip
\author{Hartmuth Arenh\"ovel$^{1}$, Winfried Leidemann$^{2}$,
and Edward L. Tomusiak$^{3}$\\
$^{1}$Institut f\"ur Kernphysik,
Johannes Gutenberg-Universit\"at,\\
 D-55099 Mainz, Germany\\
$^{2}$Dipartimento di Fisica, Universit\`a di Trento, and\\
Istituto Nazionale di Fisica Nucleare, Gruppo collegato di Trento,\\
 I-38050 Povo, Italy\\
$^{3}$Department of Physics and Engineering Physics\\
and Saskatchewan Accelerator Laboratory,\\
University of Saskatchewan, Saskatoon, Canada S7N 0W0
}
\maketitle
\begin{abstract}
For deuteron photo- and electrodisintegration the selection of complete 
sets of polarization observables is discussed in detail by applying a 
recently developed new criterion for the check of completeness of a chosen 
set of observables. The question of ambiguities and their resolution by 
considering additional observables is discussed for a numerical example, 
for which the role of experimental uncertainties is also investigated. 
Furthermore, by inversion of the expressions of the observables as 
hermitean forms in the $t$-matrix elements a bilinear term of the form 
$t_{j'}^*t_j$ can be given as a complex linear form in the observables 
from which an explicit solution for $t_j$ in terms of observables can 
be obtained. These can also be used to select sets of observables for 
the explicit representation of the $t$-matrix. 
\end{abstract}
\section{Introduction}
Complete information on the dynamics of a reaction is contained in its 
reaction or $t$-matrix. Thus its determination is most desirable
in order to provide a basis  for a detailed comparison with any theoretical 
model. In general, the reaction matrix elements are complex numbers for which 
one phase remains undetermined, i.e.\ arbitrary. 
This means, if ones has $n$ independent $t$-matrix elements, one needs at least
$2n-1$ independent observables for a complete 
determination of all matrix elements, 
although in general some discrete ambiguities remain with such a minimal 
number of observables. On the other hand, since the number of linearly 
independent observables is $n^2$, the question arises of how one can decide 
whether a given set of $2n-1$ polarization observables of a reaction 
constitutes a complete set from which the reaction matrix can be reconstructed.
In a recent paper \cite{ALT98} we have derived a general criterion which 
allows one to decide this question unambiguously. 
In that work we also have illustrated this criterion by applying 
it to the longitudinal observables of deuteron electrodisintegration. 

In the present work, we have extended this analysis to all polarization 
observables of electromagnetic deuteron break-up, i.e., photo- and 
electrodisintegration. To this end we briefly review in Sect.\ 
\ref{formalism} our previous work on the general formalism of polarization 
observables \cite{Are88,ArS90,ALT93}, for which we had chosen a particular 
basis for the representation of the $t$-matrix elements, and generalize 
it to arbitrary orientations 
of the initial and final spin quantization axes. Sect.\ \ref{criterion} is 
devoted to the application of our new criterion to the transverse 
polarization observables, deriving explicitly groups 
of possible complete sets. We will also discuss for a specific numerical 
example the discrete ambiguities which can appear for a complete set with 
a minimal number of observables. Furthermore, we will analyze in this section
the influence 
of possible experimental errors in such an analysis and derive bounds, 
which specific polarization observables have to fulfil. 
In Sect.\ \ref{bilin} we will show that an analytic 
solution of the reaction matrix elements in terms of structure 
functions is possible. This is achieved by inverting the formal 
expressions of the structure functions as hermitean forms in the reaction 
matrix elements yielding the bilinear terms $t_{j'}^*t_j$ as linear 
superpositions of structure functions. Their explicit form 
depends on the chosen basis for the initial and final hadronic spin states. 
Several choices will be discussed in detail. We will close with a summary and
outlook. Specific details and complimentary material are presented in several 
appendices. In particular, in Appendix A a detailed comparison between our 
formalism and the one of Dmitrasinovic and Gross \cite{DmG89} is given. 

\section{General Formalism}\label{formalism}
In this section we will briefly review the general formalism of polarization
observables in deuteron photo- and electrodisintegration - the latter in the 
one-photon-approximation - as presented in detail in \cite{ArS90,ALT93} 
with some generalization to arbitrary quantization axes for the initial and 
final hadronic states. 

We start from Eq.\ (58) of \cite{ALT93} for a general observable $X$ in 
exclusive deuteron electrodisintegration
\beqa
{\cal O}(X)&=&P(X) S_0\nonumber\\
&=&c(k_1^{lab},k^{lab}_2) \sum _{I=0}^2 P_I^d \sum _{M=0}^I
 \Big\{(\rho _L f_L^{IM}(X) + \rho_T f_T^{IM}(X) + 
\rho_{LT} {f}_{LT}^{IM+}(X) \cos \phi \nonumber\\
& &\qquad \qquad \qquad + \rho _{TT} {f}_{TT}^{IM+}(X) \cos2 \phi)
\cos (M\tilde{\phi}-\bar\delta_{I}^{X} {\pi \over 2}) \nonumber\\
& &\;\;-(\rho_{LT} {f}_{LT}^{IM-}(X) \sin \phi 
+ \rho _{TT} {f}_{TT}^{IM-}(X) \sin2 \phi) 
\sin (M\tilde{\phi}-\bar\delta_{I}^{X} {\pi \over 2})\nonumber\\
& &\mbox{}+ h \Big[ (\rho'_T f_T^{\prime\, IM}(X) 
+ \rho '_{LT} {f}_{LT}^{\prime\, IM-}(X) \cos \phi )
\sin (M\tilde{\phi}-\bar\delta_{I}^{X} {\pi \over 2}) \nonumber\\
& &\qquad + \rho '_{LT} {f}_{LT}^{\prime\, IM+}(X) \sin \phi 
\cos (M\tilde{\phi}-\bar\delta_{I}^{X} {\pi \over 2})\Big] 
\Big\} d_{M0}^I(\theta_d)\,,\label{obsfin}
\eeqa
where $S_0$ denotes the unpolarized differential cross section and 
\begin{equation}
c(k_1^{lab},k^{lab}_2) = {\alpha \over 6 \pi^2} 
                          {k^{lab}_2 \over k_1^{lab} q_{\nu}^4}\;.
\end{equation}
Here $\alpha$ is the fine structure constant and 
$k_1 ^{lab}$ and $k_2 ^{lab}$ denote the
lab frame momenta of the initial and the scattered electrons, respectively, 
while $q_{\nu} ^2=q_0^2-\vec q^{\,2}$ is the squared four momentum transfer 
$(q = k_1 - k_2).$
The virtual photon density matrix is given by
\beqa
\rho_L = -\beta^2 q^2_{\nu} {\xi^2 \over 2 \eta}\,,\hspace{.9cm}& &\hspace{1cm}
\rho_{LT} = -\beta q^2_{\nu} {\xi \over \eta} \sqrt{{\xi + \eta \over 8}}\;,
\\
\rho_T = -{1 \over 2} q^2_{\nu} (1 + {\xi \over 2 \eta})\,,& &\hspace{1cm}
\rho_{TT} =  q^2_{\nu} {\xi \over 4 \eta}\;,\\
\rho'_{T} =- {1 \over 2} q^2_{\nu} \sqrt{{\xi + \eta \over \eta}}
\,,\,\,\,\,& &\hspace{1cm}
\rho'_{LT} = -{1 \over 2} \beta q^2_{\nu} {\xi \over \sqrt{2 \eta}}
\;,
\eeqa
with
\begin{equation}
\beta = {|{\vec q}^{\,lab}| \over |{\vec q}^{\,c}|},\,\,\,\,\,
\xi = -{q^2_{\nu} \over ({\vec q}^{\,lab})^{\,2}},\,\,\,\,\,
\eta = {\rm tan}^2({\theta_e^{lab} \over 2})\;,
\end{equation}
where $\theta_e^{lab}$ denotes the electron scattering angle in the lab system 
and $\beta$ expresses the boost from the lab system to the frame in which the 
hadronic tensor is evaluated and ${\vec q}^{\,c}$ denotes the momentum 
transfer in this frame. 
For real photons only the transverse structure functions contribute and 
one has to replace in (\ref{obsfin}) 
$c(k_1^{lab},k^{lab}_2)\rightarrow 1/3$ and the virtual 
photon density matrix by
\beqa
\matrix{\rho_L \rightarrow 0,\,&\rho_{LT} \rightarrow 0,\,&
\rho'_{LT} \rightarrow 0,\,\cr
\rho_T \rightarrow \frac{1}{2},\,&
h\rho'_T \rightarrow \frac{1}{2}P^\gamma_c,\,& 
\rho_{TT} \rightarrow -\frac{1}{2}P^\gamma_l\,,\cr}
\eeqa
where $P^\gamma_l$ and $P^\gamma_c$ denote the degree of linear and circular
photon polarization, respectively. 

\begin{figure}
\centerline{%
\epsfxsize=12.0cm
\epsffile{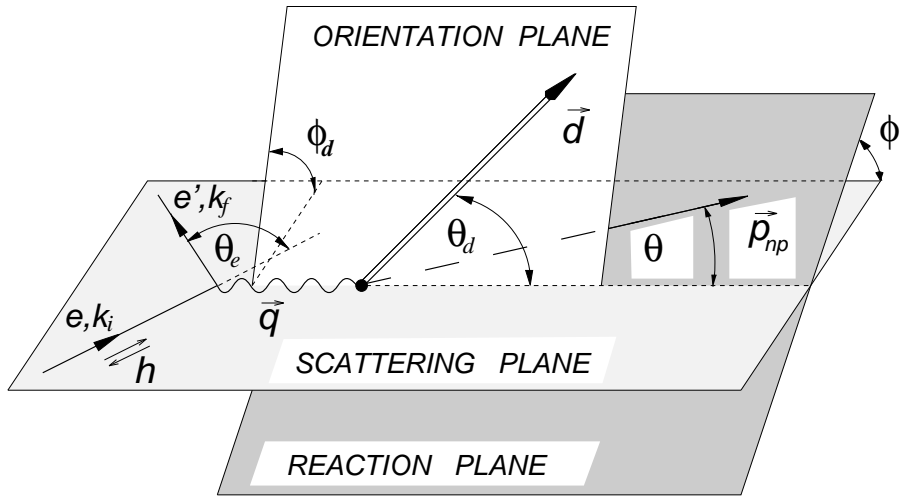}
}
\vspace*{1cm}
\caption{Geometry of exclusive electron-deuteron scattering with
polarized electrons and an oriented deuteron target. The relative $np$ 
momentum, denoted by \protect{${\vec p}_{np}$}, is characterized by angles 
\protect{$\theta=\theta_{np}$} and \protect{$\phi=\phi_{np}$} where the 
deuteron orientation axis, denoted by \protect{$\vec d$}, is specified 
by angles \protect{$\theta_d$} and \protect{$\phi_d$}.
\label{fig1}
}
\end{figure}

Often the hadronic tensor and thus 
the observables are calculated in the final $np$ c.m.\ system. This 
system, which sometimes is also called anti-lab system, moves in the 
laboratory with total momentum ${\vec q}^{\,lab}$. In the following we will
adopt this system, i.e.\ ${\vec q}^{\,c}={\vec q}^{\,c.m.}$. 
The initial state is characterized by the photon helicity 
$\lambda$ ($=\pm 1$ for real and $=0,\,\pm 1$ for virtual photons),  
and the deuteron spin projection $\lambda_d$ 
with respect to a chosen quantization 
axis. Correspondingly, the final $np$ system is characterized by the relative 
$np$ momentum $\vec p_{np}$ and the spin quantum numbers of the two nucleons, 
either in the uncoupled representation $(\lambda_p,\lambda_n)$ or the 
coupled one $(s,m_s)$ with respect to some quantization axis. 
The deuteron polarization is described by spherical orientation tensors 
$\tau^{[I]}_M$ with $I=0,1,2$, and for the outgoing nucleon polarization 
components including no polarization one has 
16 independent operators acting in the final 
two-nucleon spin space according to all combinations of the four operators 
$(1,\,\vec \sigma)$ of each of the two nucleons. The components of 
the spin operators of both particles refer to the reference frame associated 
with the final $np$ c.m.\ system denoted by $(x,\,y,\,z)$ where the 
$z$-axis is parallel to 
$\vec p_{np}$ in the reaction plane and its $y$-axis parallel 
to $\vec q\times \vec p_{np}$, i.e.\ perpendicular to the reaction plane. Thus 
the polarization components of particle ``1'' (here the proton) 
are chosen according to the Madison convention while for particle ``2'' 
(neutron) the $y$- and $z$-components 
of $\vec P$ have to be reversed in order to comply with this convention. 
The spherical angles of proton and neutron momenta 
with respect to the reference 
frame associated with the reaction plane in the c.m. system are 
$\theta^{c.m.}_p=\theta$, $\phi^{c.m.}_p=\phi$ and 
$\theta^{c.m.}_n=\pi-\theta$, $\phi^{c.m.}_n=\phi+\pi$ 
(see Fig.\ \ref{fig1}).

The basic quantities which determine the differential cross section and 
the outgoing nucleon polarization components for beam and target polarization
and which contain the complete information 
on the dynamical properties of the $NN$ system available in deuteron 
photo- and electrodisintegration  
are the structure functions $f^{(\prime) I M}_\alpha (X)$, where $\alpha 
\in \{L,T,LT,TT\}$ characterizes the diagonal and interference 
contributions from the longitudinal and transverse polarizations of the 
virtual or real photon, and $(I,M)$ the deuteron polarization tensors. 
The polarization components of the outgoing two nucleons are represented 
by $X=(x_{\alpha'}x_\alpha)$ where the first entry refers to the proton 
and the second to the neutron. Explicitly we use $x_1=x$, $x_2=y$, $x_3=z$, 
whereas $x_0=1$ describes the case of no polarization. Furthermore, for 
the explicit notation of a polarization component of only one nucleon we 
use an index 1 or 2 in case of proton or neutron, respectively, i.e., 
instead of e.g.\ $X=(x1)$ we use $X=x_1$ or instead of $X=(1z)$ we use 
$X=z_2$. An explicit listing is given in Table~\ref{tab1} where we also have  
indicated the division of the observables into two 
sets, named $A$ and $B$, according to their behaviour under a parity 
transformation as discussed in~\cite{ArS90}.
Formal expressions of structure functions have also been derived in 
\cite{DmG89} with respect to the socalled hybrid basis. The relation of 
these structure functions to ours is established in Appendix A. 

In \cite{ALT93} we have expressed all structure functions in terms of real or 
imaginary parts of the quantities ${\cal U}_X^{\lambda' \lambda I M}$ 
\beqa
f_{L}^{IM}(X)&=&\frac{2}{1+\delta_{M0}}\Re e\left(i^{\bar \delta^X_I}
{\cal U}^{00 I M}_{X}\right)\,,\label{fL}\\
f_{T}^{IM}(X)&=&\frac{4}{1+\delta_{M0}}\Re e\left(i^{\bar \delta^X_I}
{\cal U}^{11 I M}_{X}\right)\,,\\
f_{LT}^{IM\pm}(X)&=&\frac{4}{1+\delta_{M0}}\Re e\left[i^{\bar \delta^X_I}
\left({\cal U}^{01 I M}_{X}\pm(-)^{I+M+\delta_{X,\,B}}
{\cal U}^{01 I -M}_{X}\right)\right]\,,\\
f_{TT}^{IM\pm}(X)&=&\frac{2}{1+\delta_{M0}}\Re e\left[i^{\bar \delta^X_I}
\left({\cal U}^{-11 I M}_{X}\pm(-)^{I+M+\delta_{X,\,B}}
{\cal U}^{-11 I -M}_{X}\right)\right]\,,\\
f_{T}^{\prime IM}(X)&=&\frac{4}{1+\delta_{M0}}\Re e\left(i^{1+\bar \delta^X_I}
{\cal U}^{11 I M}_{X}\right)\,,\\
f_{LT}^{\prime IM\pm}(X)&=&\frac{4}{1+\delta_{M0}}\Re e
\left[i^{1+\bar \delta^X_I}
\left({\cal U}^{01 I M}_{X}\pm(-)^{I+M+\delta_{X,\,B}}
{\cal U}^{01 I -M}_{X}\right)\right]\,.\label{fLTp}
\eeqa
Here we have introduced
\beq
\bar \delta_I^X:= (\delta_{X,B}-\delta_{I1})^2\,, \mbox{ and }
\delta_{X,B}:=\left\{\matrix{1 & \mbox{for}\; X\in B\cr 0 & 
\mbox{for}\; X\in A\cr} \right.\,,
\eeq
distinguishing the two sets of observables $A$ and $B$. 

The total number of structure functions listed in (\ref{fL}) through 
(\ref{fLTp}) is $2n^2$ \cite{ALT93}, where the number $n$ of independent 
matrix elements is 12 in photo- and 18 in electrodisintegration. This 
is twice as much as the number of linearly independent observables. Indeed, 
one finds $n^2$ linear relations among the structure functions 
(see \cite{ArS90} for photo- and \cite{ALT93} for electrodisintegration),
reducing the total number of linearly independent observables to the 
required $n^2$. 
On the other hand, since each reaction matrix element is in general a complex 
number, but one overall phase is undetermined, a set of $2n-1$ properly 
chosen observables should suffice to determine completely all 
matrix elements. This seeming contradiction is resolved by the 
observation, that the linearly independent observables are 
not completely independent of each other in a more general sense. In fact, 
since any bilinear form $t_{j'}^\ast t_j$ can be given as a linear 
expression in the observables (see Sect.\ IV), one can find exactly 
$(n-1)^2$ quadratic relations between them as is shown in Appendix B,
reducing the total number of independent observables just to 
the required number. Consequently, one can determine all matrix elements from
$2n-1$ properly chosen observables. However, one should keep in mind that 
the solution is in general not unique but contains discrete ambiguities. 

Now we will proceed to review the explicit representation of the structure
functions as quadratic hermitean forms in the $t$-matrix elements. In the 
foregoing expressions, the ${\cal U}$'s are given as bilinear forms in 
the reaction matrix elements, i.e., for $X=(x_{\alpha'}x_\alpha)$
\beq
{\cal U}_{\alpha'\alpha}^{\lambda' \lambda I M}= 
\sum_{m_1'm_2'\lambda_d' m_1 m_2 \lambda_d}
t^*_{m_1'm_2'\lambda'\lambda_d'}\langle m_1'm_2'|
\Omega_{\alpha'}(1)\Omega_{\alpha}(2)|m_1 m_2\rangle
t_{m_1 m_2 \lambda \lambda_d}\langle \lambda_d|\tau^{[I]}_M|\lambda_d'\rangle
\,,\label{ulamcart}
\eeq
where $(m_1,m_2)$ stands for the spin quantum numbers of the outgoing 
nucleons, either in the coupled representation $(s,m_s)$ or in the 
uncoupled one $(\lambda_p,\lambda_n)$, and the spin operators for the 
outgoing nucleons are denoted by
\beq
\Omega_{\alpha}(i)=\sigma_{\alpha}(i)\,,\quad (i=1,2),
\eeq
with $\alpha=0,\dots,3$ and $\sigma_0={\mathbb 1}_2$. 
In view of the angular momentum algebra it is useful 
to switch to a spherical representation replacing the cartesian 
components of the spin operators by their spherical ones  
\beqa
\Omega_{\alpha}(i)&=&\sum_{\tau=0,1}\sum_{\nu=-\tau}^{\tau}
   s_{\alpha}^{\tau\nu}\,\Omega_{\tau\nu}(i)\,,\\ 
\Omega_{\tau\nu}(i)&=&\sigma_{\nu}^{[\tau]}(i)
\,\qquad (\tau = 0,\,1),\label{sphcom}
\eeqa
defining $\sigma^{[0]}={\mathbb 1}_2$ and $\sigma^{[1]}_\nu=\sigma_\nu$. 
The transformation matrix is defined by 
\beq
s_{\alpha}^{\tau\nu}=\bar c(\alpha)\,\delta_{\tau \widetilde \tau(\alpha)}\,
  (\delta_{\nu\widetilde\nu (\alpha)}+
   \hat c(\alpha)\,\delta_{\nu -\widetilde\nu (\alpha)})\,,
\eeq
with 
\beq
\begin{array}{ll}
\hat c(\alpha) = \delta_{\alpha 2} - \delta_{\alpha 1}\,, & 
\bar c(\alpha) = 
  \left\{\matrix{1 & \mbox{for }\alpha=0,\,3\cr
      \frac{i^{-\alpha-1}}{\sqrt{2}} & \mbox{for }\alpha=1,\,2\cr}
\right. \,,\\ & \\
\widetilde \tau (\alpha) = 1- \delta_{\alpha 0}\,, &
\widetilde \nu (\alpha) =   
  \left\{\matrix{0 & \mbox{for }\alpha=0,\,3\cr
       1 & \mbox{for }\alpha=1,\,2\cr}
\right.\,.\end{array}
\eeq
Explicit expressions are listed in Table \ref{tabcartsph}.
For later purposes we note the inverse transformation from spherical to 
cartesian components
\beq
\Omega_{\tau\nu}(i)=\sum_{\alpha=0}^3 c^\alpha_{\tau\nu}\,\Omega_{\alpha}(i)\,,
\label{sphcart}
\eeq
where we have defined 
\beq
c_{\tau \nu}^{\alpha} = c(\nu)
       (\delta_{\alpha a(\tau, \nu)}
 +i\nu\,\delta_{\alpha b(\tau, \nu)})\,,
\eeq
with
\beq
c\,(\nu) = -\frac{\nu}{\sqrt{2}}\,\delta_{|\nu| 1}+\delta_{\nu 0}\,,\quad
a\,(\tau, \nu) = 3\tau - 2|\nu|\,,\quad
b\,(\tau, \nu) = 3\tau - |\nu|\,.
\eeq
Then the transformation of the ${\cal U}$'s to spherical components is 
given by
\beq
{\cal U}_{\alpha'\alpha}^{\lambda' \lambda I M}=
\sum_{\tau'\nu'\tau\nu}s_{\alpha'}^{\tau'\nu'}s_{\alpha}^{\tau\nu}\,
{\cal U}^{\lambda' \lambda I M}
_{\tau'\nu'\tau\nu}\,
\label{cartsph}
\eeq
where ${\cal U}_{\tau'\nu'\tau\nu}^{\lambda' \lambda I M}$ is defined as in
(\ref{ulamcart}) with $\Omega_{\alpha}(i)$ being replaced by the spherical
components of (\ref{sphcom}).

The explicit form in terms of the $t$-matrix elements 
depends on the representation for the initial and final 
spin states. 
As mentioned above, we will consider two cases, an uncoupled representation
with spin quantum numbers $(\lambda_p,\, \lambda_n)$ and a coupled one with 
$(s,\, m_s)$. The canonical direction of the quantization 
axis for the initial deuteron state is the incoming photon momentum and 
for the final state the direction of the relative $np$ momentum in the 
final state c.m.\ frame. In the uncoupled representation this is called the 
helicity representation and for the coupled case we have named it the 
standard representation. 
The representations obey the following symmetry relation if parity is 
conserved
\beqa
t_{-\lambda_p -\lambda_n -\lambda -\lambda_d}&=&
(-)^{\lambda_p +\lambda_n+ \lambda+ \lambda_d}\,
t_{\lambda_p \lambda_n \lambda \lambda_d}\,,\label{symhel}\\
t_{s -m_s -\lambda -\lambda_d}&=&(-)^{1+s+ m_s+ \lambda+ \lambda_d}\,
t_{s m_s \lambda \lambda_d}\,.
\label{symst}
\eeqa
The transformation from one representation to the other is simply given by a 
Clebsch-Gordan coefficient
\beq
t_{\lambda_p \lambda_n \lambda \lambda_d}=\sum_{s m_s} (-)^{m_s}\hat s
 \left(\matrix{\frac{1}{2} & \frac{1}{2} & s \cr
    \lambda_p  & \lambda_n & -m_s \cr} \right) t_{s m_s \lambda \lambda_d}\,.
\eeq
A third representation called hybrid basis, where the quantization axis 
is chosen perpendicular to the reaction plane, was introduced in \cite{DmG89}. 
We shall also consider this basis later on.

For later purposes, however, it is useful to allow for arbitrary 
directions of the quantization axis. Thus we will consider general rotations
$R_d(\alpha_d,\beta_d,\gamma_d)$ of the initial deuteron state and 
$R_f(\alpha_f,\beta_f,\gamma_f)$ of the final two-nucleon state. Denoting
a rotated state by a subscript ``$R$'', i.e.\ $|jm\rangle_R$, one has 
\beqa
\langle j m|j \bar m\rangle_R &=& 
D^{j}_{m \bar m}(R)\,,
\eeqa
where the $D$-matrices are taken in the convention of \cite{Ros63}.
Correspondingly, the initial and final state irreducible spin operators 
transform under a rotation $R$ as
\beq
R^{-1}O^{[j]}_{m}R=\sum_{m'}O^{[j]}_{m'}D^{j}_{m'm}(R^{-1})\,.
\eeq
Then the reaction matrix elements are transformed according to 
\beqa
t^{R_f R_d}_{\bar \lambda_p \bar \lambda_n \lambda \bar \lambda_d} &=&
\sum_{\lambda_p \lambda_n \lambda_d}
{_{R_f}\langle}\frac{1}{2}\bar \lambda_p|\frac{1}{2} \lambda_p\rangle \,
{_{R_f}\langle}\frac{1}{2}\bar \lambda_n|\frac{1}{2} \lambda_n\rangle 
\,t_{\lambda_p \lambda_n \lambda \lambda_d}\,
\langle 1 \lambda_d|1 \bar \lambda_d\rangle_{R_d} \nonumber\\
&=&\sum_{\lambda_p \lambda_n \lambda_d}
D^{\frac{1}{2}}_{\bar \lambda_p\lambda_p}(R_f^{-1})\,
D^{\frac{1}{2}}_{\bar \lambda_n\lambda_n}(R_f^{-1})\,
\,t_{\lambda_p \lambda_n \lambda \lambda_d}\,
D^1_{\lambda_d \bar \lambda_d}(R_d)
\eeqa
for the uncoupled representation, and 
\beq
t^{R_f R_d}_{s \bar m_s \lambda \bar \lambda_d} =
\sum_{m_s \lambda_d}D^s_{\bar m_s m_s}(R_f^{-1})
\,t_{s m_s \lambda \lambda_d}\,
D^1_{\lambda_d \bar \lambda_d}(R_d)
\eeq
for the coupled one. The symmetry of (\ref{symhel}) and (\ref{symst}) 
translates into a somewhat more involved relation for the case of 
arbitrary rotations $R_f$ and $R_d$, namely one finds
\beqa
t^{R_f R_d}_{-\lambda_p -\lambda_n -\lambda -\lambda_d}&=&
(-)^{\lambda_p +\lambda_n+ \lambda+ \lambda_d}\,
t^{\bar R_f \bar R_d}_{\lambda_p \lambda_n \lambda \lambda_d}\,,
\label{symhelgen}\\
t^{R_f R_d}_{s -m_s -\lambda -\lambda_d}&=&(-)^{1+s+ m_s+ \lambda+ \lambda_d}\,
t^{\bar R_f \bar R_d}_{s m_s \lambda \lambda_d}\,,
\label{symstgen}
\eeqa
where we have defined for a rotation $R=(\alpha,\beta,\gamma)$ an associated
rotation by $\bar R=(-\alpha,\beta,-\gamma)$.
We note in passing that the transformation to the hybrid basis of \cite{DmG89}
is achieved by a simultaneous rotation around the $x$-axis by $\pi/2$, i.e.,
$R_f=R_d$ with $(\alpha,\beta,\gamma)=(\pi/2,\pi/2,-\pi/2)$. 

For the ${\cal U}$'s, the evaluation of the spin operators in (\ref{ulamcart})
leads to
\beq
{\cal U}_{\tau'\nu'\tau\nu}^{\lambda' \lambda I M}=
\sum_{m_1'm_2'\lambda_d' m_1 m_2 \lambda_d}
C^{m_1'm_2'\lambda_d'}_{m_1 m_2 \lambda_d}(\tau'\nu'\tau\nu IM)\,
(t^{R_f R_d}_{m_1'm_2'\lambda'\lambda_d'})^* \,
t^{R_f R_d}_{m_1 m_2 \lambda \lambda_d}
\,,\label{Ugeneral}
\eeq
where one has for the uncoupled representation
\beqa
C^{\lambda_p' \lambda_n' \lambda_d'}_{\lambda_p \lambda_n \lambda_d}
(\tau'\nu'\tau\nu IM)=2\sqrt{3}\, 
\Omega^{\tau'\nu'}_{\frac{1}{2}\lambda_p'\frac{1}{2}\lambda_p}(R_f^{-1})\,
\Omega^{\tau\nu}_{\frac{1}{2}\lambda_n'\frac{1}{2}\lambda_n}(R_f^{-1})\,
\Omega^{IM}_{1\lambda_d 1\lambda_d'}(R_d^{-1})\,.\label{Cuncoup}
\eeqa
Here we have introduced the quantities
\beq
\Omega^{J M}_{j'm' j m}(R)=(-)^{j'-m'} \hat J\sum_{M'}
 \left(\matrix{j' & J & j\cr -m' & M' & m \cr} \right)D^J_{M'M}(R)\,,
\eeq
for which we would like to note the following properties: \\
(i) symmetry  
\beqa
\Omega^{J M}_{j m j' m'}(R)&=&(-)^{J+m-m'}\Omega^{J M}_{j' -m' j -m}(R)\,,
\label{Omsym1}\\
(\Omega^{J M}_{j'm' j m}(R))^*&=&(-)^{j+m+j'+m'+J+M}
\Omega^{J -M}_{j'-m' j -m}(R)\,,\label{Omsym2}
\eeqa
(ii) orthogonality
\beq
\sum_{JM}(\Omega^{J M}_{j'm' j m}(R))^*\,\Omega^{J M}_{j'\bar m' j \bar m}(R)
= \delta_{\bar m' m'}\delta_{\bar m m}\,.\label{ortho}
\eeq
Correspondingly, one finds for the coupled representation
\beqa
C^{s' m_s' \lambda_d'}_{s m_s \lambda_d}(\tau'\nu'\tau\nu IM)&=& 
(-)^{\tau'+\tau}2\sqrt{3}\,\hat\tau'\hat \tau\hat s'\hat s \,
\Omega^{IM}_{1\lambda_d 1\lambda_d'}(R_d^{-1})\nonumber\\
& &\sum_{S\sigma}(-)^\sigma {\hat S}
\left( \matrix {\tau'&\tau &S \cr \nu'&\nu&-\sigma \cr}\right)
\left\{ \matrix {\frac{1}{2}&\frac{1}{2}&\tau'\cr 
\frac{1}{2}&\frac{1}{2}&\tau \cr s'&s &S \cr} \right\}
\Omega^{S\sigma}_{s'm_s' s m_s}(R_f^{-1})\,.\label{Ccoup}
\eeqa
For both cases one easily finds with the help of (\ref{ortho}) and the 
orthogonality properties of the $3j$- and $9j$-symbols the orthogonality 
relation 
\beq
\sum_{\tau'\nu'\tau\nu IM}
(C^{\bar m_1'\bar m_2'\bar \lambda_d'}_{\bar m_1 \bar m_2 \bar \lambda_d}
(\tau'\nu'\tau\nu IM))^*\,
C^{m_1'm_2'\lambda_d'}_{m_1 m_2 \lambda_d}(\tau'\nu'\tau\nu IM)\,
= 12 \delta_{\bar m_1'm_1'}\delta_{\bar m_2' m_2'}
\delta_{\bar \lambda_d'\lambda_d'}\delta_{\bar m_1 m_1}
\delta_{\bar m_2 m_2}\delta_{\bar \lambda_d \lambda_d}\,,\label{orthoC}
\eeq
where one has $(m_1,\,m_2)=(\lambda_p,\,\lambda_n)$ for the uncoupled 
case, and $(m_1,\,m_2)=(s,\,m_s)$ for the coupled case. 
It is worth mentioning, that the ${\cal U}$'s do not depend on the choice 
of representation as is apparent from the general definition in 
(\ref{ulamcart}) as a trace with respect to the spin quantum numbers, because 
it is invariant under unitary transformations of the spin states. 

The ${\cal U}$'s possess the following symmetry properties
\beq
\Big({\cal U}_X^{\lambda'\lambda IM}\Big)^*= 
(-)^{M}{\cal U}_X^{\lambda\lambda' I -M}=
(-)^{\lambda'+\lambda + I +\delta_{X,B}}
{\cal U}^{-\lambda -\lambda' I M}_{X}\,,
\label{symcart}
\eeq
or for the spherical representation
\beq
\Big({\cal U}^{\lambda'\lambda IM}_{\tau'\nu'\tau\nu}\Big)^*
=(-)^{M+\nu'+\nu}{\cal U}^{\lambda\lambda'I-M}_{\tau'-\nu'\tau-\nu}
=(-)^{\lambda'+\lambda +I+\tau'+\tau}
{\cal U}^{-\lambda-\lambda' I M}_{\tau'\nu'\tau\nu}\,.
\label{symsph}
\eeq
These relations can be proven most easily using (\ref{symhel}) or 
(\ref{symst}) in conjunction with (\ref{Omsym1})-(\ref{Omsym2}) for 
$R_f=R_d=(0,0,0)$, i.e., for the helicity or standard representation. 

\section{General Criterion for Complete sets of observables}\label{criterion}

We will now address the question whether a set of $2n-1$ observables, 
chosen from the set of $n^2$ linearly independent observables, constitutes a 
complete set. In \cite{ALT98} we have derived a general criterion which 
allows one to decide this question uniquely. Before applying it to the 
present reaction, we will give first a brief outline of the main result 
of \cite{ALT98}. 

Any observable in a reaction with $n$ independent matrix elements can be 
represented by an $n\times n$ hermitean form $f^\alpha$ 
in the complex $n$-dimensional variable $z$  
\begin{eqnarray}
f^\alpha(z)&=&\frac{1}{2}\sum_{j'j} z_{j'}^* F_{j'j}^\alpha z_j\,,\label{genf}
\end{eqnarray}
where hermiticity requires 
\begin{eqnarray}
(F_{j'j}^{\alpha})^*=F_{jj'}^\alpha\,,
\end{eqnarray}
and $z$ comprises all independent reaction matrix elements labeled by 
$j$. 

For the application of our criterion one first has to rewrite the 
hermitean form in (\ref{genf}) into a real quadratic form by introducing
\begin{eqnarray}
z&=&x+i y\,,\\
F^\alpha&=&A^\alpha+i\, B^\alpha \,,
\end{eqnarray}
where $A^\alpha$ and $B^\alpha$ are real matrices, and $A^\alpha$ is 
symmetric whereas $B^\alpha$ is antisymmetric.
Considering further the fact that one overall phase is arbitrary, one may 
choose $y_{j_0}=0$ for one index $j_0$ and then one finds for the given 
observable
\begin{eqnarray}
f^\alpha(x+iy)&=&\frac{1}{2}\Big[\sum_{j'j} x_{j'} A_{j'j}^\alpha x_j
+\sum_{\tilde j'\tilde j} y_{j'} A_{j'j}^\alpha y_j 
+2\sum_{\tilde j'j} y_{j'} B_{j'j}^\alpha x_j\Big]
\,,
\end{eqnarray}
where the tilde over a summation index indicates that the index $j_0$ has to  
be left out. Introducing now an $(m=2n-1)$-dimensional real vector $u$ by 
\begin{eqnarray}
u=( x_1, \dots , x_n , y_1, \dots , y_{j_0-1}, y_{j_0+1}, \dots , y_n)\,,
\end{eqnarray}
one can represent the $n\times n$ hermitean form by an $m\times m$
real quadratic form
\begin{eqnarray}
\widetilde f^\alpha(u)&=&\frac{1}{2}\sum_{l'l=1}^{m} u_{l'} 
                     \widetilde F_{l'l}^\alpha u_l\,,
\end{eqnarray}
where the $m\times m$ matrix $\widetilde F^\alpha$ is given by 
\begin{eqnarray}
\widetilde F^\alpha=\left(\begin{array}{cc} A^\alpha & 
(\widetilde B^\alpha)^T \\
\widetilde B^\alpha &  {\widehat A}^\alpha \end{array}\right)\,.
\end{eqnarray}
Here $\widetilde B^\alpha$ is obtained from $B^\alpha$ by canceling the 
$j_0$-th row, and ${\widehat A}^\alpha$ from $A^\alpha$ by canceling 
the $j_0$-th row and column. Thus $\widetilde B^\alpha$ is an $(n-1)\times n$ 
matrix and $ {\widehat A}^\alpha$ an $(n-1)\times(n-1)$ matrix. 

Now, for checking the completeness of a chosen set of $2n-1$ observables 
one has to construct the $m\times m$ corresponding matrices 
$\widetilde F^\alpha$, and then one builds from their columns for all possible
sets $\{k_1,\dots,k_m;\, k_\alpha \in \{1,\dots,n\}\}$ the matrices  
\begin{eqnarray}
\widetilde W(k_1,\dots,k_m)=\left(\begin{array}{ccc} 
\widetilde F^1_{1k_1} & \cdots &\widetilde  F^m_{1k_m} \\ 
\vdots & &\vdots \\ \widetilde F^1_{mk_1} & \cdots & \widetilde F^m_{mk_m} 
\end{array}\right)
\,.
\end{eqnarray}
Note that the $k_\alpha$ need not be different. If at least one of the 
determinants of $\widetilde W(k_1,\dots,k_m)$ is 
nonvanishing then one has a complete set. 

Now we will apply this criterion for the selection of complete sets to the 
case of deuteron photo- and electrodisintegration. The total number 
of $T$-matrix elements for electrodisintegration 
is $3\times 3\times 4=36$ which is reduced by parity conservation to 
$n=18$. Of these, 6 are associated with the charge or longitudinal current 
density component while the remaining 12 belong to the transverse current 
density components. Only the latter appear in photodisintegration. 
In order to apply our criterion, one first has to construct the matrices 
$\widetilde F$ which represent the structure functions as hermitean forms 
in the reaction matrix elements as
\beq
f_\alpha^{(\prime)\, I M\pm}(X)=
  \sum_{j'j} t^*_{j'}\widetilde F^{(\prime)\, I M \pm,\,\alpha}_{j'j}
  (X)\, t_j\,.\label{Fmatrix}
\eeq
According to (\ref{fL}) through (\ref{fLTp})
one needs the matrix representation of
\beq
{\cal U}^{\lambda' \lambda I M}_{X}=
  \sum_{j'j} t^*_{j'}\widetilde C^{I M \lambda' \lambda}_{j'j}
  (X) t_j\,,
\eeq
which can be read off from Eqs.\ (\ref{cartsph}) and (\ref{Ugeneral}) 
yielding with $X=(x_{\alpha'}x_\alpha)$
\beq
\widetilde C^{I M \lambda' \lambda}_{j'j}(X)=
\sum_{\tau'\nu'\tau\nu}s_{\alpha'}^{\tau'\nu'}s_{\alpha}^{\tau\nu}\,
C^{m_1'm_2'\lambda_d'}_{m_1 m_2 \lambda_d}(\tau'\nu'\tau\nu IM)\,,
\eeq
where the labeling is to be understood as $j^{(\prime)}=(m^{(\prime)}_1,
m^{(\prime)}_2, \lambda^{(\prime)}, \lambda^{(\prime)}_d)$. 
Detailed expressions of the $\widetilde C^{I M \lambda' \lambda}(X)$'s 
for several representations are easily obtained 
from the expressions listed in Appendix C. 
Then the the $\widetilde F$'s are given in terms of the $\widetilde C$'s 
as
\beqa
\widetilde F_{j'j}^{IM,\,L}(X)&=&\frac{2}{1+\delta_{M0}}i^{\bar \delta^X_I}
\widetilde C^{I M 00}_{j'j}(X)\,,\label{hL}\\
\widetilde F_{j'j}^{IM,\,T}(X)&=&\frac{2}{1+\delta_{M0}}i^{\bar \delta^X_I}
\Big[\widetilde C^{I M 11}_{j'j}(X) + (-)^{\bar \delta^X_I}
\Big( \widetilde C^{I M 11}_{j'j}(X)\Big)^*\Big]\,,\\
\widetilde F_{j'j}^{IM\pm,\,LT}(X)&=&\frac{2}{1+\delta_{M0}}i^{\bar \delta^X_I}
\Big[\Big(\widetilde C^{I M 0 1}_{j'j}(X)\pm(-)^{I+M+\delta_{X,\,B}}
\widetilde C^{I -M 01}_{j'j}(X)\Big)\nonumber\\
&&\hspace*{1cm}+ (-)^{\bar \delta^X_I}
\Big(\widetilde C^{I M 0 1}_{j'j}(X)\pm(-)^{I+M+\delta_{X,\,B}}
\widetilde C^{I -M 01}_{j'j}(X)\Big)^*\Big]\,,\\
\widetilde F_{j'j}^{IM\pm,\,TT}(X)&=&\frac{2}{1+\delta_{M0}}i^{\bar \delta^X_I}
\Big(\widetilde C^{I M -1 1}_{j'j}(X)\pm(-)^{I+M+\delta_{X,\,B}}
\widetilde C^{I -M -1 1}_{j'j}(X)\Big)\,,\\
\widetilde F_{j'j}^{\prime IM,\,T}(X)&=&\frac{2}{1+\delta_{M0}}i^{1+\bar 
\delta^X_I}
\Big[\widetilde C^{I M 1 1}_{j'j}(X)-(-)^{\bar \delta^X_I}
\Big(\widetilde C^{I M 1 1}_{j'j}(X)\Big)^*\Big]\,,\\
\widetilde F_{j'j}^{\prime IM\pm,\,LT}(X)&=&\frac{2}{1+\delta_{M0}}i^{\bar 
\delta^X_I}
\Big[\Big(\widetilde C^{I M 0 1}_{j'j}(X)\pm(-)^{I+M+\delta_{X,\,B}}
\widetilde C^{I -M 01}_{j'j}(X)\Big)\nonumber\\
&&\hspace*{1cm}- (-)^{\bar \delta^X_I}
\Big(\widetilde C^{I M 0 1}_{j'j}(X)\pm(-)^{I+M+\delta_{X,\,B}}
\widetilde C^{I -M 01}_{j'j}(X)\Big)^*\Big]\,.
\eeqa
It is convenient to arrange the labeling of the $t$-matrix elements in 
such a way that the longitudinal ones belong to $j=1,\dots,6$ and the 
transverse ones to $j=7,\dots,18$. Thus the general structure of these 
matrices is then 
\beq
\widetilde F^\alpha = 
   \left(\begin{array}{cc}
   A^\alpha & C^\alpha\\
   (C^\alpha)^\dagger & B^\alpha
   \end{array}\right),\,
\eeq
where $A^\alpha$ is a $(6\times 6)$-matrix, $C^\alpha$ a 
$(6\times 12)$-matrix, and $B^\alpha$ a $(12\times 12)$-matrix. In particular
one has 
\beq
\widetilde F^L = 
   \left(\begin{array}{cc}
   A^L & 0 \\
   0 & 0
   \end{array}\right),\,
\widetilde F^{(\prime)\,T/TT} = 
   \left(\begin{array}{cc}
   0 & 0 \\
   0 & B^{(\prime)\,T/TT}
   \end{array}\right),\,\mbox{and }
\widetilde F^{(\prime)\,LT} = 
   \left(\begin{array}{cc}
   0 & C^{(\prime)\,LT} \\
   (C^{(\prime)\,LT})^\dagger & 0
   \end{array}\right).
\eeq

The structure of these matrices is such that the longitudinal ($L$) and the 
transverse ($T,\,TT$) observables are decoupled filling separated $6\times 6$-
and $12\times 12$-submatrices, respectively, whereas the $LT$-type
observables are represented by $18\times 18$-matrices. These features offer
various kinds of strategies for selecting complete sets. 
\begin{itemize}
\item[(i)]
One may independently select complete sets of observables for the 
longitudinal and transverse cases. In other words, 
one may choose a set of 11 longitudinal structure functions 
for a check of completeness, and analogously 23 transverse structure functions.
With respect to the latter, one has in view of the linear relations between 
the $T$- and the $TT^+$-type and between the $T'$- and $TT^-$-type 
observables different choices, taking either $T$- and $T'$-type or 
$TT^{\pm}$-type observables or even mixing different types of observables. 
The missing relative phase between the longitudinal and transverse 
$t$-matrix elements can then be provided by any $LT$-observable. The advantage
of this approach is that in this way one automatically obtains complete 
sets of observables for the case of photodisintegration as well, namely the 
transverse ones. 
\item[(ii)]
Again one may start with a selection of 11 longitudinal structure functions 
yielding the longitudinal $t$-matrix elements. 
But then instead of choosing transverse observables, one may directly choose
24 linearly independent $LT$-type observables which then constitute a simple 
system of linear equations for the missing transverse matrix elements, 
because the longitudinal ones are then known from the first step. 
\item[(iii)]
Complimentary to case (ii) one may start with a selection of 23 transverse
observables taking one of the alternatives listed in (i). Then a proper set of 
12 $LT$-type observables should provide a set of linear equations from which
the missing longitudinal $t$-matrix elements can be obtained. 
\item[(iv)]
An alternative to the foregoing procedures would be a selection of 
35 structure functions of $LT$-type. However, in this case the completeness 
check would be much more involved due to the considerably higher dimension 
of the determinants to be checked. 
\end{itemize}
The question which of these strategies is more advantageous will depend on
the experimental conditions. Often $L$- and $T$-type structure functions 
are easier to determine in an experiment although the required Rosenbluth 
separation introduces some unwanted complication. In view of the fact that 
the strategies (i) through (iii) require the determination of either $L$- 
or $T$-type observables or both we will consider exclusively in the 
following analysis the question of complete sets for longitudinal and 
transverse structure functions. 

The longitudinal case has already been discussed in detail in \cite{ALT98}. 
We will briefly summarize the main result. For the analysis of possible 
complete sets we had chosen the helicity basis. Taking into account the 
symmetry
property of the $t$-matrix we have used for the labeling of the six 
independent longitudinal matrix elements the ones listed in Table 
\ref{tabhelL}. Allowing general choices for $(k_1,\dots,k_{11})$, we found 
that there is only a very weak restriction on the possible
sets. In fact one may select any set, which does not include more than
eight observables of the type $X^{10}$ and $X^{22}$ of the set of linearly 
independent observables chosen. They were listed explicitly for $X\in 
\{1,xx,xz,y_1,x_1,x_2,z_1,z_2\}$ in Table 5 of \cite{ALT98}. As mentioned in 
\cite{ALT98}, this is most easily seen by looking at the structure of the 
matrices $\widetilde F^{(\prime)\, I M \pm,\,\alpha}(X)$ associated with each 
observable (see (\ref{Fmatrix})) in the helicity basis. But this feature is 
independent of the basis chosen for the representation of the $t$-matrices.

For the transverse observables one can proceed in an analogous way. The 
independent helicity matrix elements in the helicity basis are listed 
in Table \ref{tabhelT}. Again with respect to the general question of a 
complete set of observables one finds the general statement, that one 
may pick any set of 23 
observables from the chosen set of linearly independent ones with the only 
restriction, that not more than 16 should be of the type $X^{(\prime)\,10}$ 
and $X^{(\prime)\,22}$. 

In Ref.\ \cite{ALT98} we simulated an experimental study for the 
determination of the longitudinal $t$-matrix elements of the helicity basis 
from a given set of ``measured'' observables whose numerical values 
were taken from a calculation. Various complete sets were
selected and the arising system of 11 nonlinear equations for the
$t$-matrix elements was solved. Since the solutions are not
unique we had to calculate additional observables, henceforth called 
``check observables'' , taking as input the obtained 
solutions for the $t$-matrix elements and compare them to their ``measured''
values. For the arbitrarily chosen kinematics (internal excitation energy
$E_{np}=100$ MeV, momentum transfer $q^2_{c.m.}=5$ fm$^{-2}$, various
$np$ angles $\theta_{np}$) we found that one of the considered complete
sets was particularly suitable (first set of Table 6 in Ref. \cite{ALT98}). 
In this case only one additional check
observable ($f^{10}_L(x_2)$) was sufficient to determine the correct solution. 
In the present work we have extended this simulation to a somewhat more 
realistic experimental situation using the same kinematics ($\theta_{np}=50$
degrees) again and taking the same specific set but allowing for errors in the 
measured observables. The size of the error is chosen in two different ways 
(cases A and B). 

For case A we have assumed a statistical error of 10 \% for any
of the eleven observables (standard deviation $\sigma(f_L^{IM}(X)) = 
0.1 f_L^{IM}(X))$. We varied each of the 11 observables randomly, subject 
to the restriction that the standard deviation of each observable was as 
given above. A statistically correct randomization procedure was used.
We note in passing that in varying the observables one has to pay 
attention to the fact that they 
have to fulfill certain boundary conditions (see Appendix D).
With these randomly assigned values we have solved again the system of 
equations and then calculated from the solutions the check observable. 
If the value of the check 
observable did not differ more than 10 \% from its measured value the
solution was accepted. We repeated this procedure 10$^5$ times obtaining
about 1500 accepted solutions. For these successful solutions we have 
calculated the mean values and standard deviations (experimental errors) 
for the 11 real and imaginary parts of the $t$-matrix elements. We had 
chosen $\Im m (t_1)=0$. 
The results are shown in Tables \ref{tab1exp} and \ref{tab2exp}. 
Though the mean values come quite close
to the correct values of the $t$-matrix elements (average deviation 
about 10 \%)
the standard deviations from the mean values are on the average 44 \% and
thus the experimental error would be quite large. We have tried to improve
our results by considering in addition a second check observable 
($f^{22}_L(y_1)$). In this case the number of
accepted solutions is reduced to about 400 and the results were much
better. In fact we found an average deviation from the true value of 6.5 \%.
In addition the experimental error came down to an average value of 25 \%  
which, however, is still quite sizable.

One may argue that case A is not very realistic, since we had
assumed the same relative error for all observables independent of their size. 
However, observables with large values can probably be measured with a higher 
precision than observables with small values. Thus for case B we made a 
different choice. We assumed that the largest observable can be determined 
with a relative error of 1 \% and then we take the resulting absolute 
uncertainty for all the other observables as their absolute error. 
For our smallest observable this led to a relative error of about 40 \%.
With one check observable we found about 1300 accepted solutions and obtained
average values very close to the true values (average deviation 1.0 \%), but 
the experimental error is on the average 22 \% and thus still 
quite large. Considering two check observables for case B also leads to 
an important improvement. With the 440 successful solutions
we found an average experimental error of only 4.0 \%. The
average deviation of the mean
values from the exact values is similar as before (1.2 \%). 

Our simulation of an experimental situation shows that one can get rather 
reliable results for the $t$-matrix elements even if experimental errors 
are taken into account. The results can be greatly improved if 
additional check observables are considered.
For our case it was sufficient to consider two such observables. 
If on the other hand one uses no check observables at all 
one gets rather unreliable results since other types of solutions of
the nonlinear system of equations are mixed in. 
In fact, performing our simulation
without any check observable leads to large experimental errors 
(average error more than 100 \%) and also to strong 
average deviations of the mean values from the true values of the $t$-matrix
elements (about 50 \%).  

We also made a similar study for the determination of the
transverse $t$-matrix elements from observables, but without introducing 
experimental errors. However, the transverse case is much more complicated than
the longitudinal one due to the higher dimensionality (12 instead of 
6 complex $t$-matrix elements), since one has to solve a system of 23 nonlinear
equations. We have used two different complete sets, which
are listed in Table \ref{tabexpT}. The first set has been chosen arbitrarily
considering only target asymmetries and observables with proton polarization 
components $P_x$ and $P_y$. On the contrary, the second set was chosen 
using one of the sets obtained from the inversion of the bilinear
$t$-matrix expressions as discussed in the next section. In both cases we took
4 additional check observables, namely $f_T^{10}(x_2)$, $f_T^{11}(x_2)$,
$f^{\prime\,00}_T(x_2)$, and $f_T^{11}(z_2)$ for the first set and  
$f_T^{22}(1)$, $f^{\prime\,00}_T(x_1)$, $f_T^{10}(x_2)$, and 
$f_T^{11}(x_2)$ for the second set. Our numerical method to solve the 
system of equations asks for starting values for the 23 real and 
imaginary parts of the $t$-matrix elements. We have determined these 
starting values on a random basis. For both complete sets we made $10^6$ 
trials. In order to classify the solutions we have evaluated the check
observables from the solution. Then we summed the squares of their 
differences to their true values and took it as a measure for the 
quality of the solution. The best solutions
of the first set led to rather good results for the real parts of the
$t$-matrix elements, while the smaller imaginary parts were not so well
described. Thus no correct solution had been found with $10^6$ trials.
Of course one could increase the number of trials or search for a more
efficient choice for the starting values. For the second complete
set the situation was much better. Though we did not find a completely
correct solution with $10^6$ trials, the best solution came extremely close
to the correct solution. Ten of the 23 real and imaginary parts of the 
$t$-matrix elements were determined
better than 1 \% and most of the rest also with a rather good relative
precision. Larger deviations were only found in three cases, but these
concern extremely small matrix elements which are $10^2$ - $10^3$ times smaller
than the largest $t$-matrix element. Thus the solution could be considered 
to be practically correct, and this shows that in principle the method 
works also for the transverse case.

\section{Derivation of bilinear $t$-matrix expressions}\label{bilin}

One can also derive a direct solution of the reaction matrix 
elements in terms of observables, because one can 
express all bilinear forms $t_{j'}^*t_j$ as linear 
combinations of the structure functions $f^{(\prime) I M}_\alpha (X)$, i.e.,
\beq
t_{j'}^*t_j=T_{j'j}[f^{(\prime) I M}_\alpha (X)]\,.
\eeq
This is possible 
because the spin operators representing the various polarization degrees 
of freedom form a complete basis of operators in spin space. 
An explicit solution has been reported in \cite{ArS90} for the case of 
deuteron photodisintegration. In order to extend this case to 
electrodisintegration, we invert first the relations in (\ref{fL}) through 
(\ref{fLTp}) yielding the following expressions for $M\ge 0$ in 
\beqa
{\cal U}_X^{00\, I M} &=& \frac{1}{2}(1+\delta_{M0})\, i^{-\bar \delta_I^X} 
                   f_L^{IM}(X)\,,\\
{\cal U}_X^{11\, I M} &=& \frac{1}{4}(1+\delta_{M0})\, i^{-\bar \delta_I^X} 
                   \Big(f_T^{IM}(X) - i\,f_T^{\prime\,IM}(X)\Big) \,,\\
{\cal U}_X^{01\, I M} &=& \frac{1}{8}(1+\delta_{M0})\, i^{-\bar \delta_I^X} 
                \Big(f_{LT}^{IM+}(X) + f_{LT}^{IM-}(X) 
        - i\,(f_{LT}^{\prime\,IM+}(X) + f_{LT}^{\prime\,IM+}(X))\Big) \,,\\
{\cal U}_X^{01\, I -M} &=& \frac{(-)^M}{8}(1+\delta_{M0})\, 
                i^{\bar \delta_I^X} \Big(f_{LT}^{IM+}(X) - f_{LT}^{IM-}(X) 
        - i\,(f_{LT}^{\prime\,IM+}(X) - f_{LT}^{\prime\,IM+}(X))\Big) \,,\\
{\cal U}_X^{-11\, I M} &=& \frac{1}{4}(1+\delta_{M0})\, i^{-\bar \delta_I^X} 
                \Big(f_{TT}^{IM+}(X) + f_{TT}^{IM-}(X)\Big) \,,\\
{\cal U}_X^{-11\, I -M} &=& \frac{(-)^M}{4}(1+\delta_{M0})\, 
             i^{\bar \delta_I^X}\Big(f_{TT}^{IM+}(X) - f_{TT}^{IM-}(X)\Big) \,.
\eeqa
The other ${\cal U}$'s not listed above can be obtained from the 
foregoing ones by use of the symmetry relations in (\ref{symcart}). 

The ${\cal U}_X^{\lambda' \lambda I M}$'s 
in turn are given as linear forms in $t_{j'}^*t_j$ 
of the reduced $t$-matrix elements  
which one can invert so that any $t_{j'}^*t_j$ can be expressed as linear 
superposition of the ${\cal U}_X^{\lambda \lambda' I M}$'s 
\beqa
t_{j'}^{R_f R_d\,*}t_j^{R_f R_d} = \sum_{IM} \sum_{\alpha', \alpha=0}^3 
              \widetilde T_{j' j}^{IM \alpha' \alpha}\,
              {\cal U}_{\alpha'\alpha}^{\lambda' \lambda I M}\,.
\label{ttobs}
\eeqa
The inversion is easily achieved. Starting from (\ref{cartsph}) and 
(\ref{Ugeneral}) 
and using the orthogonality relation of (\ref{orthoC}) in conjunction 
with (\ref{sphcart}), one obtains for 
the general representation ($j^{(\prime)}=(m^{(\prime)}_1,
m^{(\prime)}_2, \lambda^{(\prime)}, \lambda^{(\prime)}_d)$)
\beq
\widetilde T_{j' j}^{IM \alpha' \alpha}=\frac{1}{12}\sum_{\tau'\nu'\tau\nu}
(C^{m_1'm_2'\lambda_d'}_{m_1 m_2 \lambda_d}(\tau'\nu'\tau\nu IM))^*
\,c_{\tau' \nu'}^{\alpha'}\,c_{\tau \nu}^{\alpha}\,,\label{tij}
\eeq
where for the uncoupled basis $(m_1,\,m_2)=(\lambda_p,\,\lambda_n)$, 
and the matrix elements are labeled by $j^{(\prime)}=(\lambda^{(\prime)}_p,
\lambda^{(\prime)}_n, \lambda^{(\prime)}, \lambda^{(\prime)}_d)$, whereas 
for the coupled basis $(m_1,\,m_2)=(s,\,m_s)$ and $j^{(\prime)}=
(s^{(\prime)},m^{(\prime)}_s, \lambda^{(\prime)}, m^{(\prime)}_d)$.
More explicit expressions for the normal helicity, hybrid and standard 
bases can be obtained with the help of Appendix C. 

The bilinear relations in (\ref{ttobs}) can now be exploited in various ways. 
One possibility is to choose a fixed matrix element, say $t_{j_0}$, as real
and positive. Then all other matrix elements $t_j$ with $j\neq j_0$ are 
uniquely determined relative to $t_{j_0}$ and given as linear forms of
appropriate structure functions \cite{ArS90}
\beq
t_j = \frac{1}{t_{j_0}} T_{j_0j}[f^{(\prime)IM\pm}_\alpha(X)]\,.\label{t_j}
\eeq
Finally, for the determination of the missing matrix element $t_{j_0}$ one
has to choose only one additional structure function, say $f_0$, which has
the general form (see (\ref{Fmatrix}))
\beq
f_0=\sum_{j'j}t_{j'}^* \tilde F_{j'j}t_j\,.\label{f_0}
\eeq
Inserting $t_j$ from (\ref{t_j}) one finds 
\beq
t_{j_0}=\sqrt{\frac{\sum_{j'j}T_{j'j_0}\tilde F_{j'j} T_{j_0 j}}
                   {f_0}}\,.\label{tj0}
\eeq
However, proceeding in this way, one needs in general a much larger number of 
observables for the complete determination of the $t$-matrix 
than the required minimal number of $2n-1$ of a complete set of a 
$n$-dimensional $t$-matrix. 

A more general strategy which leads in general to a smaller number 
of necessary observables is to study first all interference terms 
with respect to the question, which and how many observables are involved, 
because a closer inspection of the explicit expressions reveals, that in 
general they can be divided into subgroups which are governed by a 
restricted number of observables. In order to visualize this grouping 
we have devised a graphical representation. To this 
end we assemble the numbers ``1'' through ``$n$'' 
by points on a circle and represent 
a term $t_{j'}^*t_j$ by a straight line joining the points ``$j$'' and 
``$j'$''. Interference terms belonging to the same group are then 
represented by the same type of lines as is demonstrated below. 

As next step one has to choose from the total number of interference 
terms $t_{j'}^*t_j$  with $j'>j$ which is $\frac{1}{2}n(n-1)$ -- not 
counting $t_{j'}^*t_j$ with $j'<j$, because $(t_{j'}^*t_j)^*=t_{j}^*t_{j'}$ 
-- a set of $n-1$ independent interference terms. The meaning of ``independent 
interference terms'' can be explained most easily by a graphical 
representation. Assembling again the numbers as before, we consider 
first a set of connected interference terms, which means that they 
generate a pattern of connected lines so that any point belonging to one of 
the considered interference terms is connected to any other point of the 
set either directly or via $k$ other points of that set. In such a set 
any matrix element $t_{j'}$ can be expressed in terms of any other matrix 
element $t_j$ of that set by one of the two forms
\beq
t_{j'}=\left\{\begin{array}{ll} \frac{T_{j_1 j'}}{T_{j_1 j_2}}
\frac{T_{j_3 j_2}}{T_{j_3 j_4}}\cdots 
\frac{T_{j_{k-2} j_{k-3}}}{T_{j_{k-2} j_{k-1}}}
\frac{T_{j_k j_{k-1}}}{T_{j_k j}}\,t_j & \mbox{ for }k\,\mbox{ odd},\\
\frac{T_{j_1 j'}}{T_{j_1 j_2}}
\frac{T_{j_3 j_2}}{T_{j_3 j_4}}\cdots 
\frac{T_{j_{k-3} j_{k-4}}}{T_{j_{k-3} j_{k-2}}}
\frac{T_{j_{k-1} j_{k-2}}}{T_{j_{k-1} j_k}}\frac{T_{j j_k}}{t_j^*} 
& \mbox{ for }k\,\mbox{ even},\end{array}\right.\label{connect}
\eeq
depending on whether the number $k$ of intermediate points 
connecting $j'$ with $j$ via the points $j_1$ through $j_k$ is odd or even. 
The proof of these equations is easily established by considering the simplest two 
cases of connecting two points via one and two intermediate points. For one 
intermediate point $j_1$ connecting $j'$ with $j$ one finds
\beq
t_{j'}=\frac{T_{j_1 j'}}{T_{j_1 j}}\,t_j\,,
\eeq
and for two intermediate points $j_1$ and $j_2$
\beq
t_{j'}=\frac{T_{j_1 j'}}{T_{j_1 j_2}}\frac{T_{j j_2}}{t_j^*}\,.
\eeq
Iteration of these cases yields obviously (\ref{connect}). 

If one has a closed loop with an even number of points then one finds from
(\ref{connect}) for $j'=j$ and $k$ odd the following condition 
\beq
\frac{T_{j_1 j}}{T_{j_1 j_2}}\frac{T_{j_3 j_2}}{T_{j_3 j_4}}\cdots 
\frac{T_{j_{k-2} j_{k-3}}}{T_{j_{k-2} j_{k-1}}}
\frac{T_{j_k j_{k-1}}}{T_{j_k j}}=1\,,
\label{evenpoints}
\eeq
or equivalently
\beq
T_{j_k j_{k-1}}T_{j_{k-2} j_{k-3}}\cdots T_{j_3 j_2}T_{j_1 j}=
T_{j_1 j_2}T_{j_3 j_4}\cdots T_{j_{k-2} j_{k-1}}T_{j_k j}\,,\label{evenpoints1}
\eeq
which means that in such a closed loop any interference term is completely 
determined by the other remaining interference terms of that loop.  
This condition thus constitutes a relation between the participating 
observables. On the other hand for a closed loop through an odd number 
of points, all participating $t$-matrix elements are completely 
determined up to one arbitrary phase because for 
$j'=j$ and $k$ even (\ref{connect}) yields 
\beq
|t_{j}|^2=\frac{T_{j_1 j}}{T_{j_1 j_2}}
\frac{T_{j_3 j_2}}{T_{j_3 j_4}}\cdots 
\frac{T_{j_{k-3} j_{k-4}}}{T_{j_{k-3} j_{k-2}}}
          \frac{T_{j_{k-1} j_{k-2}}}{T_{j_{k-1} j_k}}\,T_{j j_k}\,.
\label{oddpoints}
\eeq
This means one may choose one matrix element of that loop as real and non-negative, 
fix its modulus according to (\ref{oddpoints}) and then all other matrix elements 
of the loop are uniquely determined. 

Now a set of $n-1$ independent interference terms is represented by a pattern of 
$n-1$ lines in such a fashion that (i) each of the $n$ points is endpoint of at 
least one line, and (ii) each point is connected to all other points not 
necessarily in a direct manner but via intermediate points. It is 
obvious that in such a pattern no closed loops can be present, because one cannot 
construct a closed loop with $n-1$ lines such that all $n$ points are connected. Then 
all matrix elements can be expressed by one arbitrarily chosen 
matrix element, say $t_{j_0}$ according to (\ref{connect}). In order to 
fix the remaining undetermined matrix element $t_{j_0}$ one has to choose 
one additional observable $f_0$. From the form (\ref{f_0}) one obtains 
in general an equation of the type 
\beq
f_0= a + b\, |t_{j_0}|^{2} + c\,|t_{j_0}|^{-2}\,,\label{f_00}
\eeq
from which $t_{j_0}$ can be obtained, although not uniquely in general. 
The ideal 
situation would be such that one finds $n-1$ independent interference terms 
each of them represented by only two observables. Because in this case one 
employs just $2n-1$ observables. On the 
other hand analyzing the grouping of observables mentioned above, one will
in general not find such a situation, either the number of observables for 
a set of $n-1$ independent interference terms is larger than $2n-2$, or 
the grouping is such, that the choice of $n-1$ independent interference terms 
involves observables which govern at least one additional interference term 
leading to one or several closed loops. 

\begin{figure}
\centerline{%
\epsfxsize=9.0cm
\epsffile{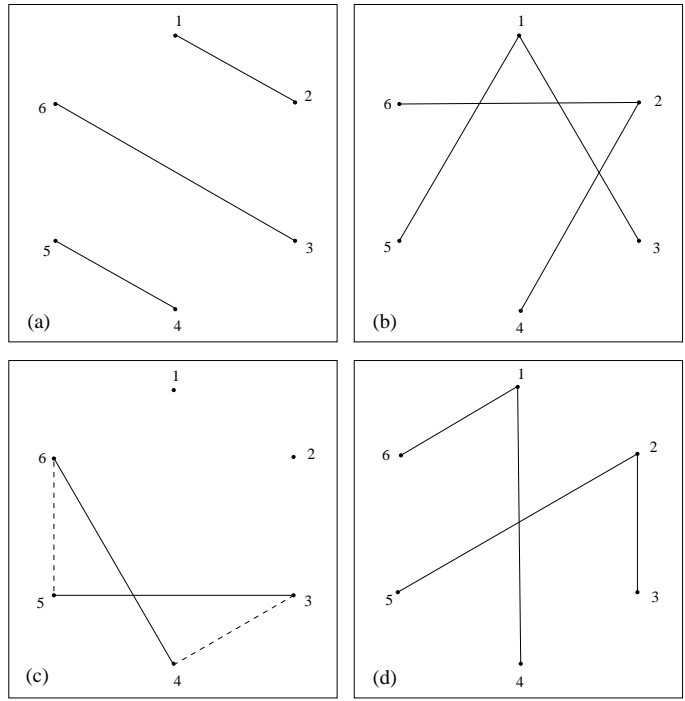}
}
\vspace*{1cm}
\caption{Diagrammatic representation of groups of longitudinal observables 
determining the interference terms of $t$-matrix elements for the helicity basis. 
The nomenclature for the groups and the corresponding observables are 
listed in Table \protect{\ref{tabgroupLhel}}.
\label{figLhel}
}
\end{figure}

We will now illustrate such an analysis for the case of the longitudinal 
matrix elements. In evaluating the 
bilinear expressions for this case, one has to use the linear relations 
between structure functions \cite{ALT93} in order to have only the linearly 
independent ones. According to Table 6 of \cite{ALT93} we have chosen for
the $A$-type observables $X= 1,y(1),xz,zz$ and for the $B$-type ones 
$X=x(1),x(2),z(1),z(2)$. We show in Fig.\ \ref{figLhel} for the helicity 
basis the graphical representation of the various groups of structure 
functions into which the interference terms divide. They are listed 
explicitly in Table \ref{tabgroupLhel}. One readily notes that 
altogether there are two groups, panel (c), each containing four observables
and each determining two different interference terms, one group (a) with six  
observables determining three interference terms of which one interference 
term, $T_{12}$, contains only four observables and finally two groups
(b) and (d) with eight observables which each determine four independent 
interference terms. 

\begin{figure}
\centerline{%
\epsfxsize=9.0cm
\epsffile{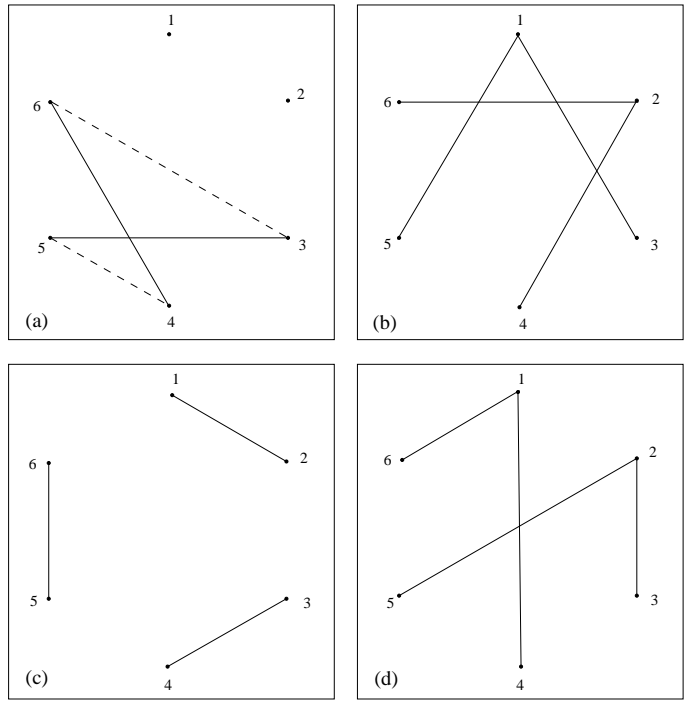}
}
\vspace*{1cm}
\caption{As Fig.\ \protect{\ref{figLhel}} for the hybrid basis with 
nomenclature in Table \protect{\ref{tabgroupLhyb}}. 
\label{figLhyb}
}
\end{figure}

The resulting grouping and nomenclature for the hybrid basis is shown 
in Fig.\ \ref{figLhyb} and Table \ref{tabgroupLhyb}, respectively. 
The pattern looks
similar to the helicity basis, 
though the number and type of observables involved are different. One finds 
in panel (a) two groups each containing six observables and each determining 
two different interference terms. Furthermore, there are three groups of eight 
observables. Two of them, (b) and (d), determine four independent
interference terms. 
They both involve the same type of single nucleon polarization components 
which differ only with respect to the particle number. The third panel (c) 
determines only three independent interference terms, of which one, $T_{21}$, 
involves only six observables. 

\begin{figure}
\centerline{%
\epsfxsize=9.0cm
\epsffile{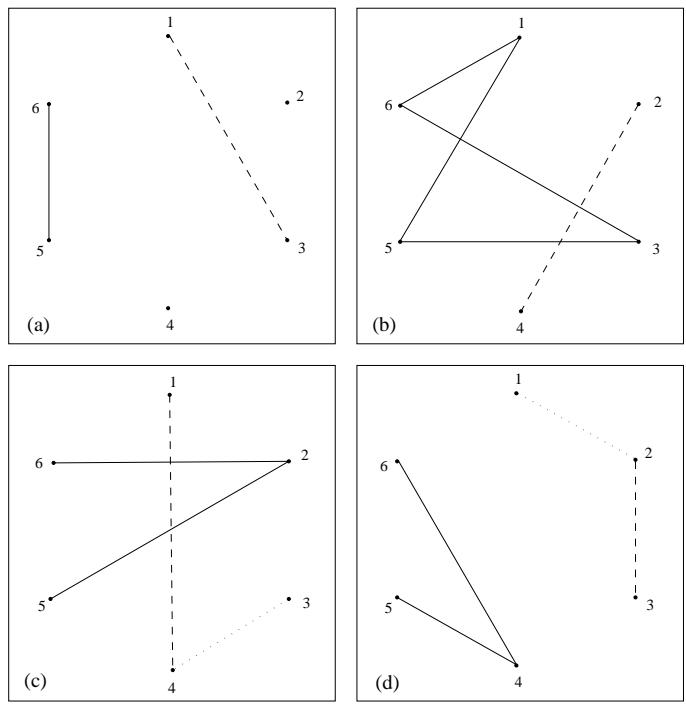}
}
\vspace*{1cm}
\caption{As Fig.\ \protect{\ref{figLhel}} for the standard  basis with 
nomenclature in Table \protect{\ref{tabgroupLstan}}. 
\label{figLstan}
}
\end{figure}

In Fig.\ \ref{figLstan} and Table \ref{tabgroupLstan} we show and explain 
the nomenclature for
 the grouping for the standard basis. The evolving pattern 
differs distinctly from the foregoing ones. First of all, one has two groups of four, 
each determining only one interference term (panel (a)). In panel (b), we show a group
of ten observables determining five interference terms containing a closed loop of four 
points and a disconnected line, i.e., a disconnected interference term. The latter is 
given by a subgroup of four observables as is indicated by the dashed line. 
Then there are two groups of eight observables for four interference terms in the panels 
(c) and (d). In each of these groups there are two connected interference terms 
which need all eight observables. Disconnected from these two, one notes two other 
connected interference terms, each of which is determined by disjunct groups of four, 
indicated by dashed and dotted lines. 

Other possible groups of observables are found by studying various rotations of the 
initial and final spin states. We 
have chosen rotations carrying the quantization axis into the $x$-axis 
($R_x=R(0,\pi/2,0)$) and the $y$-axis ($R_y=R(\pi/2,\pi/2,-\pi/2)$). We found
that rotating only the final uncoupled spin states into the $y$-axis, i.e., 
$R_f=R_y$ and $R_d=R(0,0,0)$, was most interesting. We show this case  
as our last example in Fig.\ \ref{figLhel4} supplemented by Table \ref{tabgroupLhel4}.
The pattern looks again like the one for the 
helicity basis.  However, the grouping for the transformed basis is quite different. 
One finds six groups of four observables, each determining two interference terms, and 
one 
group of six observables for three independent interference terms, 
of which one, $T_{12}$, is 
given by four observables alone. 

These patterns can now be used to select sets of observables for a complete determination 
of the $t$-matrix elements. Let us first consider the case of the helicity basis. 
It is obvious that one should take one of the groups of eight observables in panels 
(b) or (d) of Fig.\ \ref{figLhel} which fix four matrix elements in terms of two. 
Then one can choose as additional group of four from panel (c) the dashed line group 
because, taking the full line group of panel (c), 
one would obtain two disconnected 3-point loops.
Thus twelve observables lead to five complex $t$-matrix 
elements. Furthermore, since such a choice leads to a closed loop 
of an even number of points (6), one has an additional complex condition between the 
observables, eliminating two and thus leaving ten observables. In order to fix the last 
matrix element which can be chosen real and non-negative, one needs only one additional 
observable. The solution, however, could still contain an ambiguity (see (\ref{f_00})).

The largest variety of possible selections of observables is offered by the case for 
the rotated basis in Fig.\ \ref{figLhel4}. First one may take such combinations of two
of the groups of four observables determining four independent interference terms 
so that one avoids any closed loops. Then four matrix elements are given as function 
of the remaining two involving eight observables. The missing fifth interference term 
can be provided by choosing one of the remaining three groups of four observables with 
the only restriction that one should avoid two disconnected closed loops of three points. 
Obviously, one will then encounter loops of four or six points so that of the twelve 
observables involved two can be eliminated leaving one with ten observables. 

Taking into account the one remaining required observable, say $f_L$, one obtains in 
this way various groups of eleven observables, which allow one to determine the 
longitudinal matrix elements up to discrete ambiguities whose resolution needs 
additional observables \cite{ALT98}. Alternatively one could use for the missing fifth 
interference term the one containing six observables in Fig.\ \ref{figLhel4}. In that 
case one could find two closed loops resembling two conditions between observables. 
The possible scenarios are discussed explicitly in the Appendix E where we also
display the various cases diagrammatically in Figs.\ \ref{4loop} and \ref{6loop}.  

The transverse case is more involved due to the larger number of $t$-matrix elements. 
For this reason, we will not display the grouping of observables for 
all four bases as we did for the longitudinal 
matrix elements, but consider only two of them, the helicity basis and the transformed 
one. As independent matrix elements we have chosen those with $\lambda=1$ 
for the 
photon helicity. The matrix elements are numbered $7,\dots,18$ as listed in Table 
\ref{tabhelT} of Appendix E. The resulting diagrammatic representation of the 
grouping of observables is shown in Fig.\ 
\ref{figThel}. We find for both cases identical patterns for the various groups 
although with different observables (see Tables \ref{tabgroupThel} and 
\ref{tabgroupThel4} of Appendix E). In both cases we find groups of eight 
observables determining four 
interference terms (dashed lines), of twelve observables determining six interference 
terms (dash-dot-dot lines) and of sixteen observables for eight terms (solid lines). 
Combined properly according to the types of observables, they form highly symmetrical 
patterns. In panel (a) one has four disconnected loops of three points, whereas in 
the other three panels one notices a separation into two disconnected groups of 
six points building one closed loop of six points and containing in addition various 
interconnections. One then has to find proper combinations of these various groups 
such that one can express all matrix elements in terms of one. For the transverse matrix 
elements this task is more involved than the longitudinal case. A detailed analysis, 
however, shows that it is possible to find a variety of patterns containing 36 
observables, which allow a unique determination of all matrix elements as a function of 
an arbitrarily chosen one. Moreover, the number of observables can be reduced to 22, the
minimal number required, by exploiting the conditions implied by the various closed 
loops appearing in the diagrams, This will be discussed in greater detail for one case 
in the Appendix E.

\section{Summary and Conclusion}\label{summary}

In this work we have discussed various strategies for the selection of 
complete sets of polarization observables to be used for a complete 
determination of all 12, respectively 18 independent $t$-matrix elements 
for photo- and electrodisintegration of the deuteron. Such sets consist 
of 23, respectively 35 observables. Two different methods have been considered:
\begin{itemize}
\item[(i)]
Application of a newly developed criterion allowing a check of whether an 
arbitrarily chosen set of observables is complete, and 
\item[(ii)]
construction of an explicit solution 
for the $t$-matrix elements as function of observables by explicit inversion 
of the hermitean forms by which the observables are given in terms of the 
$t$-matrix elements.  
\end{itemize}
The first method is certainly much more versatile in so far as one has a much 
greater variety of choices whereby a choice may be governed by the 
question of easy access of the corresponding observables in an experiment. 
In a numerical simulation we have studied the practical applicability and 
found quite satisfactory results if a few additional check observables are 
considered which allow one to eliminate the inherent discrete ambiguities. 
But the second method has its merits, too, because it may allow an explicit 
analytic solution depending on the choice of basis for the representation of 
the $t$-matrix.
\section*{Acknowledgment}
Parts of this work were done while H.A.\ and E.L.T.\  were at the
Dipartimento di Fisica of the Universit\a`a degli Studi di Trento
and while H.A.\ and W.L.\ were visiting the
Saskatchewan Accelerator Laboratory and the Department of Physics of the
University of Saskatchewan, Saskatoon.
In all cases the authors thank the respective institutions
for their hospitality.

\setcounter{equation}{0}
\section*{Appendix A: Comparison with the Formalism of Dmitrasinovic and Gross}
\label{DG}
For the comparison with the formalism of Dmitrasinovic and Gross 
\cite{DmG89}, which henceforth we will refer to by DG, we rewrite our 
expression for the general observable in (\ref{obsfin}) as follows
\beqa
{\cal O}(X)= c(k_1^{lab},k^{lab}_2) 
\Big\{&&\rho_L\tilde f_L(X) +\rho_T \tilde f_T(X) 
 + \rho_{LT}\Big(\tilde f^{+}_{LT}(X) \cos\phi 
-\tilde f^{-}_{LT}(X) \sin\phi\Big)
\nonumber\\
&& +\rho_{TT}\Big(\tilde f^{+}_{TT}(X) \cos 2\phi
-\tilde f^{-}_{TT}(X) \sin 2\phi\Big)
\nonumber\\
&& +h \Big[\rho^{\prime}_T \tilde f^{\prime}_{T}(X)+ \rho^{\prime}_{LT}
\Big(\tilde f^{\prime -}_{LT}(X) \cos\phi +\tilde f^{\prime +}_{LT}(X) 
\sin\phi\Big)\Big]\Big\}\,,\label{OX}
\eeqa
where we have introduced
\beqa
\tilde f_\alpha^{(\prime) \pm}(X)=\sum_{I=0}^2 P^d_I\sum_{M=0}^I
d_{M0}^I(\theta_d)\,f_\alpha^{(\prime) IM\,\pm}(X)
\left\{\matrix{
\cos (M\tilde{\phi}-\bar\delta_{I}^{X} {\pi \over 2}) 
& \mbox{for}\;\, + \cr 
 \sin (M\tilde{\phi}-\bar\delta_{I}^{X} {\pi \over 2})\, & 
\mbox{for}\;\, - \cr} \right.\,,
\label{ftilde}
\eeqa
with the understanding for $\alpha =L$ and $T$ 
\beqa
\tilde f_{L/T}^{+}(X) &=& \tilde f_{L/T}(X)\,,\quad
\tilde f_{L/T}^{-}(X) = 0\,,\\
\tilde f_{T}^{\prime -}(X) &=& \tilde f_{T}^{\prime} (X)\,,\quad
\tilde f_{T}^{\prime +}(X) = 0\,.
\eeqa

On the other hand, DG give the general coincidence cross section including
target polarization and single outgoing nucleon polarization in the form 
(see Eq.\ (30) of DG)
\beqa
\frac{d^3\sigma}{dk_2^{lab}d\Omega_e^{lab}d\Omega_{np}^{c.m.}}&=&
\frac{\sigma_M p_{np}}{4\pi M_d}
\Big\{\Big(\frac{W}{M_d}\Big)^2v_LR_L +v_T R_T 
+\frac{W}{M_d} v_{TL}\Big(R^{(I)}_{LT} \cos\phi +R^{(II)}_{LT} \sin\phi\Big)
\nonumber\\
&& +v_{TT}\Big(R^{(I)}_{TT} \cos 2\phi+R^{(II)}_{TT} \sin 2\phi\Big)
+h \Big[v^{\prime}_T R_{T}^{\prime} + \frac{W}{M_d} v^{\prime}_{TL}
\Big(R^{\prime\,(II)}_{LT}\cos\phi 
+R^{\prime\,(I)}_{LT}\sin\phi\Big)\Big]\Big\}
\label{diffDG}
\eeqa
where we have changed slightly the original notation of DG for $\alpha=T,\,LT$
and $C=I,\,II$ by defining
\beq
R^{\prime\,(C)}_\alpha:= R^{(C)}_{\alpha'}\,.
\eeq
In contrast to DG, here $h$ denotes not only the electron helicity but 
also the degree of electron polarization, i.e.\ $|h|\le 1$. 
Furthermore, $\sigma_M$ denotes the Mott cross section, and the kinematic 
functions $v_\alpha^{(\prime)}$ are related to the $\rho_\alpha^{(\prime)}$ by
\beqa
\Big(\frac{W}{M_d}\Big)^2v_L &=& \frac{2\eta}{Q^2} \rho_L, \quad Q^2=-q^2\,,
\label{vL}\\
v_\alpha^{(\prime)} &=& \frac{2\eta}{Q^2} \rho_\alpha^{(\prime)}\,, 
\quad \alpha= T,\,TT\,,\\
\frac{W}{M_d} v_{TL}^{(\prime)} &=& 
-\frac{2\eta}{Q^2} \rho_{LT}^{(\prime)}\,.\label{vLTp}
\eeqa
Here $\beta = W/M_d$ expresses the boost from the final state c.m.\ system, 
having the invariant mass $W$, to the laboratory system, because the $R$'s 
are evaluated in the c.m.\ system. The observables  
are divided into two classes (denoted by $I$ and $II$) and given in the form 
for $\widetilde I\in\{I,II\}$ and with a changed notation from DG's 
$R^{(\widetilde I)}_{\alpha(\prime)}(P_j,T_i)$ (see Eq.\ (79) of DG) to 
$R^{(\prime)\,(\widetilde I)}_{\alpha}(X_j,i)$ 
\beq
R^{(\prime)\,(\widetilde I)}_{\alpha}=\sum_{j\in \widetilde I}P_{X_j}
R^{(\prime)\,(\widetilde I)}_{\alpha}(X_j)\,,\label{RalphaC}
\eeq
with
\beq
R^{(\prime)\,(\widetilde I)}_{\alpha}(X_j)=\sum_{i\in \widetilde I}T_i 
R^{(\prime)\,(\widetilde I)}_{\alpha}(X_j,i)\,,\label{RalphaCX}
\eeq
where $T_i$ denotes the deuteron polarization parameters in the 
so-called hybrid basis. According to the tables X through XII of DG, for 
the observables corresponding to $X=U$ and $X=P_n$ (notation of DG), which 
are of A-type, one has for class $I$ and $i=1,\dots,5$ 
\beqa
T_i \in \Big\{U,\,\sqrt{\frac{3}{2}}\,T_{10},\, \frac{1}{\sqrt{2}}\,T_{20},\,
\sqrt{3}\,\Re e (T_{22}),\,\sqrt{3}\,\Im m (T_{22})\Big\}\,,\nonumber
\eeqa
and for class $II$ and $i=6,\dots,9$ 
\beqa
T_i \in \Big\{\sqrt{\frac{3}{2}}\,\Re e (T_{11}),\,
\sqrt{\frac{3}{2}}\,\Im m (T_{11}),\,
\sqrt{\frac{3}{2}}\,\Re e (T_{21}),\,
\sqrt{\frac{3}{2}}\,\Im m (T_{21})
\Big\}\,,\nonumber
\eeqa
while for $X=P_s$ and $X=P_l$, which are of B-type, one has the opposite 
assignment of $T_i$ for classes $I$ and $II$. 
 
Inserting now (\ref{vL}) through (\ref{vLTp}) and (\ref{RalphaC}) into 
(\ref{diffDG}), one finds the correspondence of (\ref{obsfin}) in terms of 
the $R^{(\prime)\,(I/II)}_\alpha(X_j)$
\beqa
{\cal O}(X)=c(k_1^{lab},k^{lab}_2) \bar c
\Big\{&&\rho_LR_L(X) +\rho_T R_T(X) 
 - \rho_{LT}\Big(R^{(I)}_{LT}(X) \cos\phi +R^{(II)}_{LT}(X) \sin\phi\Big)
\nonumber\\
&& +\rho_{TT}\Big(R^{(I)}_{TT} (X)\cos 2\phi+R^{(II)}_{TT}(X) \sin 2\phi\Big)
\nonumber\\
&& +h \Big[\rho^{\prime}_T R_{T}^{\prime}(X) - \rho^{\prime}_{LT}
\Big(R^{\prime\,(II)}_{LT}(X) \cos\phi +R^{\prime\,(I)}_{LT}(X) 
\sin\phi\Big)\Big]\Big\}\,,
\label{diffDGa}
\eeqa
where
\beq
\bar c = \frac{3\alpha \pi p_{np}}{M_d}\,.
\eeq
Comparison with our expression in (\ref{OX}) then yields the 
following correspondence
\beq
\bar c \,R^{(\prime)\,(I/II)}_\alpha(X)=s^{(\prime)\,(I/II)}_\alpha
\tilde f^{(\prime)\,\pm}_\alpha(X)\,,
\label{relR-f}
\eeq
where $s^{(I)}_{LT}=s^{(II)}_{TT}=s^{\prime\,(I/II)}_{LT}=-1$ and 
$s^{(\prime)\,(C)}_\alpha=1$ otherwise.  

Now we will proceed to find the relation between the structure functions 
$R^{(\prime)\,(I/II)}_{\alpha}(X_j,i)$ of DG and our 
$f_\alpha^{(\prime) IM\,\pm}(X)$. 
To this end, we will introduce first a more compact notation for the $R$'s. 
According to what has been said above, one finds explicitly for observables 
of A-type in class $I$ and of B-type in class $II$, i.e.\ $X\in A,\, C=I$ 
and $X\in B,\, C=II$
\beqa
\sum_{i\in C}T_i R^{(\prime)\,(C)}_{\alpha}(X,i)&=& 
R^{(\prime)\,(C)}_{\alpha}(X,1) 
+ \sqrt{\frac{3}{2}}\,T_{10}\, R^{(\prime)\,(C)}_{\alpha}(X,2) 
+ \frac{1}{\sqrt{2}}\,T_{20} \,R^{(\prime)\,(C)}_{\alpha}(X,3)\nonumber\\
&&
+ \sqrt{3} \,\Re e (T_{22}) \,R^{(\prime)\,(C)}_{\alpha}(X,4)
+ \sqrt{3} \,\Im m (T_{22})\, R^{(\prime)\,(C)}_{\alpha}(X,5)\,,
\label{R-AI}
\eeqa
and analogously for observables of A-type in class $II$ and of B-type in 
class $I$, i.e.\ $X\in A,\, C=II$ and $X\in B,\, C=I$
\beqa
\sum_{i\in C}T_i R^{(\prime)\,(C)}_{\alpha}(X,i)&=& 
\sqrt{\frac{3}{2}}\Big(\Re e (T_{11})\, R^{(\prime)\,(C)}_{\alpha}(X,6)
+ \Im m (T_{11})\, R^{(\prime)\,(C)}_{\alpha}(X,7)\nonumber\\
&&
+ \Re e (T_{21}) \,R^{(\prime)\,(C)}_{\alpha}(X,8)
+ \Im m (T_{21}) \,R^{(\prime)\,(C)}_{\alpha}(X,9)\Big)\,.
\label{R-AII}
\eeqa
Noting the property
\beq
(T_{IM})^*=(-)^M T_{I-M}\,,
\eeq
one can rewrite (\ref{R-AI}) into the form
\beq
\sum_{i\in C}T_i R^{(\prime)\,(C)}_{\alpha}(X,i)=\sum_{IM}\frac{1}{2}
(1+(-)^M) T_{IM} R^{(\prime)\,IM (C)}_\alpha(X)\,,
\eeq
where we have defined
\beqa
R^{(\prime)\,00 (C)}_\alpha(X) &=& R^{(\prime)\,(C)}_{\alpha}(X,1)\,,\\
R^{(\prime)\,10 (C)}_\alpha(X) &=& 
\sqrt{\frac{3}{2}}R^{(\prime)\,(C)}_{\alpha}(X,2)\,,\\
R^{(\prime)\,20 (C)}_\alpha(X) &=& 
\frac{1}{\sqrt{2}} R^{(\prime)\,(C)}_{\alpha}(X,3)\,,\\
R^{(\prime)\,22 (C)}_\alpha(X) &=& \frac{\sqrt{3}}{2} 
\Big(R^{(\prime)\,(C)}_{\alpha}(X,4)
-i R^{(\prime)\,(C)}_{\alpha}(X,5)\Big)\,,\\
R^{(\prime)\,2-2 (C)}_\alpha(X) &=& 
(R^{(\prime)\,22 (C)}_\alpha(X))^*\,,\\
\eeqa
and correspondingly for (\ref{R-AII})
\beq
\sum_{i\in C}T_i \,R^{(\prime)\,(C)}_{\alpha}(X,i)=\sum_{IM}\frac{1}{2}
(1-(-)^M) \,T_{IM} \,R^{(\prime)\,IM (C)}_\alpha(X)\,,
\eeq
with
\beqa
R^{(\prime)\,11 (C)}_\alpha(X) &=& \frac{1}{2}\sqrt{\frac{3}{2}} 
\Big(R^{(\prime)\,(C)}_{\alpha}(X,6)
-i R^{(\prime)\,(C)}_{\alpha}(X,7)\Big)\,,\\
R^{(\prime)\,1-1 (C)}_\alpha(X) &=& -(R^{(\prime)\,11 (C)}_\alpha(X))^*\,,\\
R^{(\prime)\,21 (C)}_\alpha(X) &=& \frac{1}{2}\sqrt{\frac{3}{2}} 
\Big(R^{(\prime)\,(C)}_{\alpha}(X,8)
-i R^{(\prime)\,(C)}_{\alpha}(X,9)\Big)\,,\\
R^{(\prime)\,2-1 (C)}_\alpha(X) &=& -(R^{(\prime)\,21 (C)}_\alpha(X))^*\,.
\eeqa
For later purposes we note the property
\beq
(R^{(\prime)\,IM (C)}_\alpha(X))^* = (-)^M R^{(\prime)\,I-M (C)}_\alpha(X)\,.
\eeq
Furthermore, for the comparison with our $f$'s it is more convenient to 
use instead of ``$I$'' and ``$II$'' for the two classes the notation ``$+$''
and ``$-$'', respectively. This then results in the compact relation
\beqa
R^{(\prime)\,\pm}_\alpha(X) &=& \sum_i\, T_i\, 
R_\alpha^{(\prime)\,(I/II)}(X,i)\nonumber\\
&=& \sum_{IM} \frac{1}{2} (1\pm (-)^{M+\delta_{X,B}})\, 
T_{IM} \,R^{(\prime)\,IM\,\pm}_\alpha(X)\,,
\label{RalphaX}
\eeqa
and the relation in (\ref{relR-f}) reads as
\beq
\bar c R^{(\prime)\,\pm}_\alpha(X)=s^{(\prime)\,\pm}_\alpha
\tilde f^{(\prime)\pm}_\alpha(X)\,,
\label{relR-fnew}
\eeq
where $s^{(\prime)\,\pm}_{\alpha}$ is analogously defined as 
$s^{(\prime)\,(C)}_{\alpha}$, i.e.\ 
$s^+_{LT}=s^-_{TT}=s^{\prime\,\pm}_{LT}=-1$ and 
$s^{(\prime)\,\pm}_\alpha=1$ otherwise. 

Next, we have to express the polarization parameters $T_{IM}$ of DG, 
which refer to the transversity basis, by our polarization parameters 
($P^d_I,\,\theta_d,\, \phi_d$). DG describe first the deuteron density 
matrix by parameters $\widetilde T_{IM}$ which refer to a coordinate system 
associated with the reaction plane having the z-axis along the deuteron 
momentum. The relation of the $T_{IM}$ in the transversity basis to the 
$\widetilde T_{IM}$ is obtained by two transformations, first one to the 
helicity basis
\beq
 \widetilde T_{IM} \longrightarrow (-)^{I+M} \widetilde T_{IM}^*
=(-)^I\widetilde T_{I-M}\,,
\eeq
and then in a second step to the transversity basis by a rotation by $-\pi/2$
around the $x$-axis
\beq
(-)^I\widetilde T_{I-M}\longrightarrow \sum_{M'}
D^{I\,*}_{MM'}(\pi/2,\pi/2,-\pi/2)
(-)^I\widetilde T_{I-M'}\,,
\eeq
with $D$-matrices in the convention of Rose \cite{Ros63}, 
giving
\beq
T_{IM} = \sum_{M'} i^{M-M'} d^I_{M M'}(\frac{\pi}{2})\, (-)^{I+M'}
\widetilde T_{IM'}^*\,.
\eeq
Furthermore, since our polarization parameters refer to a coordinate 
system associated with the scattering plane, we find
\beq
\widetilde T_{IM} = P^d_I e^{-iM\tilde \phi} d^I_{M0}(\theta_d)\,.
\eeq
Thus we have
\beq
T_{IM} = (-)^I P^d_I \sum_{M'} i^{M+M'} d^I_{M M'}(\frac{\pi}{2})
e^{iM'\tilde \phi} d^I_{M'0}(\theta_d)\,.
\eeq
Inserting this expression into (\ref{RalphaX}) and reordering it, one finds
\beqa
R^{(\prime)\,\pm}_\alpha(X) &=& \sum_{IM}(-)^I P^d_I \, A_{IM}\, 
B^{(\prime)\,IM\,\pm}_\alpha(X)\nonumber\\
&=&\sum_{I}(-)^I P^d_I \sum_{M=0}^I \frac{2}{1+\delta_{M0}}\,
\Re e(A_{IM}\,B^{(\prime)\,IM\,\pm}_\alpha(X))\,, 
\eeqa
where we have introduced
\beqa
A_{IM}&=&i^{-\bar\delta^X_I}e^{iM\tilde \phi} d^I_{M0}(\theta_d)\,,\\
B^{(\prime)\,IM\,\pm}_\alpha(X)&=& i^{M+\bar\delta^X_I}
\sum_{M'}\frac{1}{2}\,(1\pm (-)^{M'+\delta_{X,B}})
i^{M'}\,d^I_{M' M}(\frac{\pi}{2})\,R^{(\prime)\,IM'\,\pm}_\alpha(X)\,.
\eeqa
Explicitly one finds
\beqa
R^{(\prime)\,\pm}_\alpha(X) 
&=&\sum_{I}(-)^I P^d_I \sum_{M=0}^I \frac{2}{1+\delta_{M0}}\,
d^I_{M0}(\theta_d)\,
%\nonumber\\
%&&
\left\{\matrix{
\cos (M\tilde{\phi}-\bar\delta_{I}^{X} {\pi \over 2}) 
\Re e ( B^{(\prime)\,IM\,+}_\alpha(X))
& \mbox{for}\;\, + \cr 
-\sin (M\tilde{\phi}-\bar\delta_{I}^{X} {\pi \over 2})
\Im m ( B^{(\prime)\,IM\,-}_\alpha(X))\, & 
\mbox{for}\;\, - \cr} \right.
\,,
\eeqa
It is now easy to show, using $(-)^{I+\delta_{X,B}+\bar\delta^X_I}=1$, that 
\beq
(B^{(\prime)\,IM\,\pm}_\alpha(X))^*=\pm B^{(\prime)\,IM\,\pm}_\alpha(X)\,,
\eeq
which yields 
\beqa
R^{(\prime)\,\pm}_\alpha(X) 
&=&\sum_{I=0}^2(-)^I P^d_I \sum_{M=0}^I \frac{2}{1+\delta_{M0}}\,
d^I_{M0}(\theta_d)\, 
%\nonumber\\
%&&
%\hspace*{1cm}
B^{(\prime)\,IM\,\pm}_\alpha(X)
\left\{\matrix{
\cos (M\tilde{\phi}-\bar\delta_{I}^{X} {\pi \over 2}) 
& \mbox{for}\;\, + \cr 
i \sin (M\tilde{\phi}-\bar\delta_{I}^{X} {\pi \over 2})\, & 
\mbox{for}\;\, - \cr} \right.\,.
\eeqa
Comparing this expression with (\ref{relR-fnew}) in conjunction with 
(\ref{ftilde}), we finally get the desired relation
\beqa
f_\alpha^{(\prime)\,IM\,\pm}(X) &=& (-)^I i^{(1\mp1)/2+M+\bar\delta^X_I}\,
\frac{2\bar c s^{(\prime)\,\pm}_\alpha}{1+\delta_{M0}}
%\nonumber\\
%&&
\sum_{M'}\frac{1}{2}\,(1\pm (-)^{M'+\delta_{X,B}})
i^{M'}\,d^I_{M' M}(\frac{\pi}{2})\,R^{(\prime)\,IM'\,\pm}_\alpha(X)
\,.
\eeqa
The inverse reads
\beqa
R_\alpha^{(\prime)\,IM\,\pm}(X) &=& (-)^{I+M/2} i^{-(1\mp1)/2-\bar\delta^X_I}\,
\frac{s^{(\prime)\,\pm}_\alpha}{2\bar c }\,( 1\pm (-)^{M+\delta_{X,B}}) 
%\nonumber\\
%&&
\sum_{M'}\frac{1}{2}\,(1+\delta_{M'0})i^{M'}\,
d^I_{M' M}(\frac{\pi}{2})\,f^{(\prime)\,IM'\,\pm}_\alpha(X)
\,.
\eeqa

\setcounter{equation}{0}
\section*{Appendix B: Quadratic Relations between Observables}\label{quadrel}
In this appendix we will show that for a set of $n$ independent $t$-matrix 
elements $\{t_j;\,j=1\dots n\}$ one finds exactly $(n-1)^2$ quadratic 
relations between observables by which the $n^2$ linearly independent 
observables are reduced to a set of $2n-1$ independent ones. To this end 
we introduce the bilinear forms
\beq
T_{j'j}:=t_{j'}^*t_j\,,
\eeq
which can be expressed as a linear form of the observables ${\cal O}^\alpha$. 
\beq
T_{j'j}=\sum_\alpha \tau^\alpha_{j'j}\,{\cal O}^\alpha\,,
\eeq
with appropriate coefficients $\tau^\alpha_{j'j}$. They have the property
\beq
\tau^\alpha_{j'j}=\tau^{\alpha\,*}_{jj'}\,,
\eeq
which follows from $T_{j'j}=T_{jj'}^*$ and the fact that the observables 
are real quantities. 
It is straightforward to show that these bilinear forms obey the relation 
\beq
T_{j'j}T_{lm}=T_{j'm}T_{lj}\,,\label{tijlma}
\eeq
which, expressed in terms of observables, yields quadratic relations between 
the latter. In particular, choosing $k=l=m$, one finds
\beq
T_{j'j}=\frac{T_{j'k}T_{kj}}{T_{kk}}\,,\label{tijlmb}
\eeq
where $k$ can be chosen arbitrarily. It is also clear that from (\ref{tijlmb})
one can recover the relation (\ref{tijlma}). Thus we only need to consider 
the latter relation, and the question is, how many independent quadratic 
relations one can find. 

We first note, it is sufficient to consider only one specific $k$, because 
from (\ref{tijlmb}) one can derive straightforwardly the analogous relation for
any other $k'$. Second, it is sufficient to consider only the cases 
$j'\leq j$, 
because $T_{jj'}=T_{j'j}^*$. The remaining relations certainly are independent 
because of the independency of the $t$-matrix elements. Choosing then first 
$j'=j$, the case $j'=k$ yields the identity, whereas for $j'\neq k$ one finds 
\beq
T_{j'j'}T_{kk}=|T_{j'k}|^2\,,
\eeq
which constitute $(n-1)$ real quadratic relations
\beq
\sum_{\alpha \alpha'} \tau^\alpha_{j'j'}\,\tau^{\alpha'}_{kk}\,{\cal O}^\alpha 
\,{\cal O}^{\alpha'}=\sum_{\alpha \alpha'} \tau^{\alpha\,*}_{j'j}\,
\tau^{\alpha'}_{j'j}\,{\cal O}^\alpha \,{\cal O}^{\alpha'}\,.
\eeq
As next we consider the case $i<j$ for which one has $N=n(n-1)/2$ different 
pairs. Again on can discard the cases $j'=k$ or $j=k$, because they do not 
result in quadratic relations, thus ruling out $n-1$ relations. Therefore one 
finds in this case ($j'<j$) a total number of 
\beq
N-(n-1)=\frac{1}{2}(n-1)(n-2)
\eeq
different complex quadratic relations
\beq
\sum_{\alpha \alpha'} \tau^\alpha_{j'j}\,\tau^{\alpha'}_{kk}\,{\cal O}^\alpha 
\,{\cal O}^{\alpha'}=\sum_{\alpha \alpha'} \tau^{\alpha}_{j'k}\,
\tau^{\alpha'}_{kj}\,{\cal O}^\alpha \,{\cal O}^{\alpha'}\,.
\eeq
Separating these into real and imaginary parts, one finds as total number 
of independent real quadratic relations between observables
\beq
(n-1)+2\frac{1}{2}(n-1)(n-2)=(n-1)^2\,,
\eeq
which is just the required number of relations in order to reduce the 
number of linearly independent observables to $n^2-(n-1)^2=2n-1$ independent
ones.

\setcounter{equation}{0}
\section*{Appendix C: Explicit Expressions for the Matrix Representation of 
${\cal U}^{\lambda' \lambda I M}_{\tau'\nu'\tau\nu}$}

Here we list the explicit forms of the 
$C^{m_1'm_2'\lambda_d'}_{m_1 m_2 \lambda_d}(\tau'\nu'\tau\nu IM)$ of 
(\ref{Ugeneral}) for the 
helicity, hybrid and standard bases. \\
(i) For the helicity basis with labeling $(\lambda^{(\prime)}_p,
\lambda^{(\prime)}_n, \lambda^{(\prime)}, \lambda^{(\prime)}_d)$ one finds 
\beqa
C^{\lambda_p' \lambda_n' \lambda_d'}_{\lambda_p \lambda_n \lambda_d}
(\tau'\nu'\tau\nu IM)&=& 
  2\sqrt{3}(-)^{\lambda'_p+\lambda'_n -\lambda_d+\tau'+\tau} 
  \hat\tau'\hat\tau\hat I
  \left(\matrix{1 & 1 & I\cr \lambda_d' & -\lambda_d & M\cr}\right)
%\nonumber\\
%&&
%\hspace*{1cm}  
  \left(\matrix{\frac{1}{2} & \frac{1}{2} & \tau' \cr 
  \lambda_p' & -\lambda_p & \nu'\cr}\right)
  \left(\matrix{\frac{1}{2} & \frac{1}{2} & \tau \cr
       \lambda_n' & -\lambda_n & \nu\cr}\right)\,.
\eeqa
(ii) Analogously one finds in the hybrid basis with the labeling
$(\tilde\lambda^{(\prime)}_p,
\tilde\lambda^{(\prime)}_n, \lambda^{(\prime)}, \tilde\lambda^{(\prime)}_d)$, 
where for proton, neutron and deuteron the spin projections refer to the 
transverse $y$-axis, 
\beqa
C^{\tilde\lambda_p' \tilde\lambda_n' \tilde\lambda_d'}
_{\tilde\lambda_p \tilde\lambda_n \tilde\lambda_d}
(\tau'\nu'\tau\nu IM)&=&2\sqrt{3} 
  \omega^{IM}_{1\tilde\lambda_d \tilde\lambda^{\prime}_d}\,
  \Big(\omega^{\tau' \nu'}
  _{\frac{1}{2}\tilde\lambda^{\prime}_p \tilde\lambda_p}\Big)^*
  \Big(\omega^{\tau \nu}
  _{\frac{1}{2}\tilde\lambda^{\prime}_n \tilde\lambda_n}\Big)^*\,,
\eeqa
where 
\beqa
\omega^{JM}_{j m' m} &=& \Omega^{JM}_{j m'j m}(\frac{\pi}{2},-\frac{\pi}{2},
-\frac{\pi}{2})\nonumber\\
&=&(-)^{j-m'}\hat J \sum_{M'}
  \left(\matrix{j & J & j\cr -m' & M' & m\cr} 
\right)i^{M'-M}d^J_{M'M}(\frac{\pi}{2})\,.
\eeqa
(iii) Finally in the standard basis with index labeling
$(s^{(\prime)},m^{(\prime)}_s, \lambda^{(\prime)}, 
m^{(\prime)}_d)$ one has
\beqa
C^{s' m_s' \lambda_d'}_{s m_s \lambda_d}(\tau'\nu'\tau\nu IM)&=& 
  {2\sqrt{3}}(-)^{1 -m_d +s'-m'_s+\tau'+\tau}\hat s \hat s' \hat\tau'\hat\tau\,
  \left(\matrix{1 & 1 & 1\cr m_d' & -m_d & M\cr} \right)
\nonumber\\
&& 
  \sum_{S\sigma}{\hat s}^2
    \left(\matrix{\tau' & \tau & S\cr \nu' & \nu & -\sigma\cr }\right)
    \left(\matrix{s' & s & S\cr m'_s & -m_s & -\sigma\cr }\right)
\left\{\matrix{\frac{1}{2} & \frac{1}{2} & \tau' \cr
       \frac{1}{2} & \frac{1}{2} & \tau \cr
       s' & s & S\cr}\right\}\,.
\eeqa

\setcounter{equation}{0}
\section*{Appendix D: Relative Bounds for Structure Functions}
In this section we will derive relative bounds for the various structure 
functions $f_\alpha^{(\prime)IM\pm}(X)$ which are based on the following 
theorem. Given $n$ $t$-matrix elements $\{t_j;\,j=1,\dots, n\}$ and a 
hermitean matrix $\Omega_{ij}$ with eigenvalues 
$\{\lambda_j;\,j=1,\dots, n\}$, then the following relation holds
\beq
\lambda_{min}\leq \frac{ tr(t^\dagger \Omega t)}{tr(t^\dagger t)}\leq 
\lambda_{max}\,,
\eeq
where
\beq
\lambda_{min/max}= Min/Max\{\lambda_j;\,j=1,\dots, n\}\,.
\eeq

As we have shown in Sect.\ \ref{criterion}, a general structure function 
can be written in the form (see (\ref{Fmatrix}))
\beq
f_\alpha^{(\prime)IM\pm}(X)=\sum_{j'j} t_{j'}^*\,
\widetilde F_{j'j}^{(\prime)IM\pm,\,\alpha}(X)\,t_j\,,
\eeq
where $\widetilde F_{j'j}^{(\prime)IM\pm,\,\alpha}(X)$ is a hermitean 
matrix whose explicit form is listed in Sect.\ \ref{criterion}. 
Denoting by $\lambda_{\alpha, min/max}^{(\prime)IM\pm}(X)$ the minimal 
respectively maximal eigenvalue of 
$\widetilde F_{j'j}^{(\prime)IM\pm,\,\alpha}(X)$, 
one finds the following relative bounds.

(i) Longitudinal structure functions:

\beq
\frac{1}{2}\,\lambda_{L,\,min}^{IM\pm}(X)
\leq \,\frac{f_L^{IM\pm}(X)}{f_L^{00}}
\leq \, \frac{1}{2}\,\lambda_{L,\,max}^{IM\pm}(X)\,,
\eeq
because
\beq
f_L^{00}=2\,\sum_{j=1}^6t_j^*t_j\,.
\eeq

(ii) Transverse and transverse-transverse interference structure functions
($\alpha=T$ and $TT$):

\beq
\frac{1}{2}\,\lambda_{\alpha,\,min}^{(\prime)IM\pm}(X)
\leq \,\frac{f_\alpha^{(\prime)IM\pm}(X)}{f_T^{00}}
\leq \, \frac{1}{2}\,\lambda_{\alpha,\,max}^{(\prime)IM\pm}(X)\,,
\eeq
because
\beq
f_T^{00}=2\,\sum_{j=7}^{18}t_j^*t_j\,.
\eeq

(iii) Longitudinal-transverse interference structure functions:

\beq
\frac{1}{2}\,\lambda_{LT,\,min}^{(\prime)IM\pm}(X)
\leq \,\frac{f_{LT}^{(\prime)IM\pm}(X)}{f_L^{00}+f_T^{00}}
\leq \, \frac{1}{2}\,\lambda_{LT,\,max}^{(\prime)IM\pm}(X)\,,
\eeq
because
\beq
f_L^{00}+f_T^{00}=2\,\sum_{j=1}^{18}t_j^*t_j\,.
\eeq
The resulting relative bounds for the various structure functions are 
listed in the following tables.

\setcounter{equation}{0}
\section*{Appendix E: Explicit bilinear expressions} 
\label{apphel4}
In this appendix we will first consider one specific example of the longitudinal 
matrix elements using a rotated helicity basis, namely the case
where only the final state spin states are rotated into the $y$-axis. In this case 
the longitudinal $t$-matrix elements possess the symmetry
\beq
t^{R_yR_0}_{\lambda_p \lambda_n 0 -\lambda_d}= (-)^{1+\lambda_p+\lambda_n+\lambda_d}
t^{R_yR_0}_{\lambda_p \lambda_n 0 \lambda_d}\,.
\eeq
The labeling of the independent matrix elements are given in Table \ref{indLhel4}. 
One finds the following bilinear expressions where we use the notation 
$f_{L,X}^{IM}$ instead of $f_{L}^{IM}(X)$ 
\beqa
T_{1,1}&=&t^\ast_{{1\over 2} {1\over 2} 0 0}t_{{1\over 2} {1\over 2} 0 
  0} = \frac{1}{6}\Big(f_{L}^{0 0} - {\sqrt{2}}\,f_{L}^{2 0} + f_{L,\,y_1}^{0 0} - 
   {\sqrt{2}}\,f_{L,\,y_1}^{2 0}\Big) \,,\nonumber\\
T_{2,1}&=&t^\ast_{-{1\over 2} -{1\over 2} 0 0}t_{{1\over 2} {1\over 2}
   0 0} = \frac{1}{6}\Big(-f_{L,\,zz}^{0 0} + {\sqrt{2}}\,f_{L,\,zz}^{2 0}+
  i\,( f_{L,\,xz}^{0 0} - {\sqrt{2}}\,f_{L,\,xz}^{2 0} ) \Big) \,,\nonumber\\
T_{3,1}&=&t^\ast_{{1\over 2} {1\over 2} 0 1}t_{{1\over 2} 
  {1\over 2} 0 0} = \frac{1}{4\sqrt{6}}\Big(-f_{L}^{2 1} - f_{L,\,y_1}^{2 1}+
  i\,( -f_{L}^{1 1} - f_{L,\,y_1}^{1 1} ) \Big) \,,\nonumber\\
T_{4,1}&=&t^\ast_{{1\over 2} -{1\over 2} 0 1}t_{{1\over 2} 
  {1\over 2} 0 0} = \frac{1}{4\sqrt{6}}\Big(-f_{L,\,x_2}^{1 1} + f_{L,\,z_2}^{2 1}+
  i\,( -f_{L,\,x_2}^{2 1} - f_{L,\,z_2}^{1 1} ) \Big) \,,\nonumber\\
T_{5,1}&=&t^\ast_{-{1\over 2} {1\over 2} 0 1}t_{{1\over 2} 
  {1\over 2} 0 0} = \frac{1}{4\sqrt{6}}\Big(-f_{L,\,x_1}^{1 1} + f_{L,\,z_1}^{2 1}+
  i\,( -f_{L,\,x_1}^{2 1} - f_{L,\,z_1}^{1 1} ) \Big) \,,\nonumber\\
T_{6,1}&=&t^\ast_{-{1\over 2} -{1\over 2} 0 1}t_{{1\over 2}
   {1\over 2} 0 0} = \frac{1}{4\sqrt{6}}\Big(f_{L,\,xz}^{1 1} + f_{L,\,zz}^{2 1}+
  i\,( -f_{L,\,xz}^{2 1} + f_{L,\,zz}^{1 1} ) \Big) \,,\nonumber\\
T_{2,2}&=&t^\ast_{-{1\over 2} -{1\over 2} 0 0}t_{-{1\over 2} 
  -{1\over 2} 0 0} = \frac{1}{6}\Big(f_{L}^{0 0} - {\sqrt{2}}\,f_{L}^{2 0} - 
   f_{L,\,y_1}^{0 0} + {\sqrt{2}}\,f_{L,\,y_1}^{2 0}\Big) \,,\nonumber\\
T_{3,2}&=&t^\ast_{{1\over 2} {1\over 2} 0 1}t_{-{1\over 2} 
  -{1\over 2} 0 0} = \frac{1}{4\sqrt{6}}\Big(-f_{L,\,xz}^{1 1} + f_{L,\,zz}^{2 1}+
  i\,( f_{L,\,xz}^{2 1} + f_{L,\,zz}^{1 1} ) \Big) \,,\nonumber\\
T_{4,2}&=&t^\ast_{{1\over 2} -{1\over 2} 0 1}t_{-{1\over 2}
   -{1\over 2} 0 0} = \frac{1}{4\sqrt{6}}\Big(-f_{L,\,x_1}^{1 1} - f_{L,\,z_1}^{2 1}+
  i\,( -f_{L,\,x_1}^{2 1} + f_{L,\,z_1}^{1 1} ) \Big) \,,\nonumber\\
T_{5,2}&=&t^\ast_{-{1\over 2} {1\over 2} 0 1}t_{-{1\over 2}
   -{1\over 2} 0 0} = \frac{1}{4\sqrt{6}}\Big(-f_{L,\,x_2}^{1 1} - f_{L,\,z_2}^{2 1}+
  i\,( -f_{L,\,x_2}^{2 1} + f_{L,\,z_2}^{1 1} ) \Big) \,,\nonumber\\
T_{6,2}&=&t^\ast_{-{1\over 2} -{1\over 2} 0 1}t_{
  -{1\over 2} -{1\over 2} 0 0} = \frac{1}{4\sqrt{6}}\Big(-f_{L}^{2 1} + f_{L,\,y_1}^{2 1}+
  i\,( -f_{L}^{1 1} + f_{L,\,y_1}^{1 1} ) \Big) \,,\nonumber\\
T_{3,3}&=&t^\ast_{{1\over 2} {1\over 2} 0 1}t_{{1\over 2} {1\over 2}
   0 1} = \frac{1}{24}\Big(2\,f_{L}^{0 0} + {\sqrt{2}}\,f_{L}^{2 0} - 
   {\sqrt{3}}\,f_{L}^{2 2} + 2\,f_{L,\,y_1}^{0 0} %\nonumber\\
%&&\hspace{4.1cm}+ 
   +{\sqrt{2}}\,f_{L,\,y_1}^{2 0} - {\sqrt{3}}\,f_{L,\,y_1}^{2 2}\Big) \,,\nonumber\\
T_{4,3}&=&t^\ast_{{1\over 2} -{1\over 2} 0 1}t_{{1\over 2} 
  {1\over 2} 0 1} = \frac{1}{8\sqrt{3}}\Big({\sqrt{2}}\,f_{L,\,x_2}^{1 0} + f_{L,\,z_2}^{2 2}+
  i\,( -f_{L,\,x_2}^{2 2} + {\sqrt{2}}\,f_{L,\,z_2}^{1 0} ) \Big) \,,\nonumber\\
T_{5,3}&=&t^\ast_{-{1\over 2} {1\over 2} 0 1}t_{{1\over 2} 
  {1\over 2} 0 1} = \frac{1}{8\sqrt{3}}\Big({\sqrt{2}}\,f_{L,\,x_1}^{1 0} + f_{L,\,z_1}^{2 2}+
  i\,( -f_{L,\,x_1}^{2 2} + {\sqrt{2}}\,f_{L,\,z_1}^{1 0} ) \Big) \,,\nonumber\\
T_{6,3}&=&t^\ast_{-{1\over 2} -{1\over 2} 0 1}t_{{1\over 2} {1\over 2}
   0 1} = \frac{1}{24}\Big(-2\,f_{L,\,zz}^{0 0} - {\sqrt{2}}\,f_{L,\,zz}^{2 0} + 
   {\sqrt{3}}\,f_{L,\,zz}^{2 2}%\nonumber\\
%&&\hspace{4.1cm}+
 + i\,( 2\,f_{L,\,xz}^{0 0} + {\sqrt{2}}\,f_{L,\,xz}^{2 0} - 
     {\sqrt{3}}\,f_{L,\,xz}^{2 2} ) \Big) \,,\nonumber\\
T_{4,4}&=&t^\ast_{{1\over 2} -{1\over 2} 0 1}t_{{1\over 2} -{1\over 2}
   0 1} = \frac{1}{24}\Big(2\,f_{L}^{0 0} + {\sqrt{2}}\,f_{L}^{2 0} + 
   {\sqrt{3}}\,f_{L}^{2 2} + 2\,f_{L,\,y_1}^{0 0} %\nonumber\\
%&&\hspace{4.1cm}+ 
 + {\sqrt{2}}\,f_{L,\,y_1}^{2 0} + {\sqrt{3}}\,f_{L,\,y_1}^{2 2}\Big) \,,\nonumber\\
T_{5,4}&=&t^\ast_{-{1\over 2} {1\over 2} 0 1}t_{{1\over 2} -{1\over 2}
   0 1} = \frac{1}{24}\Big(2\,f_{L,\,zz}^{0 0} + {\sqrt{2}}\,f_{L,\,zz}^{2 0} + 
   {\sqrt{3}}\,f_{L,\,zz}^{2 2}%\nonumber\\
%&&\hspace{4.1cm}+
 +i\,( -2\,f_{L,\,xz}^{0 0} - {\sqrt{2}}\,f_{L,\,xz}^{2 0} - 
     {\sqrt{3}}\,f_{L,\,xz}^{2 2} ) \Big) \,,\nonumber\\
T_{6,4}&=&t^\ast_{-{1\over 2} -{1\over 2} 0 1}t_{{1\over 2}
   -{1\over 2} 0 1} = \frac{1}{8\sqrt{3}}\Big({\sqrt{2}}\,f_{L,\,x_1}^{1 0} - f_{L,\,z_1}^{2 2}+
  i\,( f_{L,\,x_1}^{2 2} + {\sqrt{2}}\,f_{L,\,z_1}^{1 0} ) \Big) \,,\nonumber\\
T_{5,5}&=&t^\ast_{-{1\over 2} {1\over 2} 0 1}t_{-{1\over 2} {1\over 2}
   0 1} = \frac{1}{24}\Big(2\,f_{L}^{0 0} + {\sqrt{2}}\,f_{L}^{2 0} + 
   {\sqrt{3}}\,f_{L}^{2 2} - 2\,f_{L,\,y_1}^{0 0} %\nonumber\\
%&&\hspace{4.1cm}- 
  -{\sqrt{2}}\,f_{L,\,y_1}^{2 0} - {\sqrt{3}}\,f_{L,\,y_1}^{2 2}\Big) \,,\nonumber\\
T_{6,5}&=&t^\ast_{-{1\over 2} -{1\over 2} 0 1}t_{
  -{1\over 2} {1\over 2} 0 1} = \frac{1}{8\sqrt{3}}\Big(
  {\sqrt{2}}\,f_{L,\,x_2}^{1 0} - f_{L,\,z_2}^{2 2}+
  i\,( f_{L,\,x_2}^{2 2} + {\sqrt{2}}\,f_{L,\,z_2}^{1 0} ) \Big) \,,\nonumber\\
T_{6,6}&=&t^\ast_{-{1\over 2} -{1\over 2} 0 1}t_{-{1\over 2} 
  -{1\over 2} 0 1} = \frac{1}{24}\Big(2\,f_{L}^{0 0} + {\sqrt{2}}\,f_{L}^{2 0} - 
   {\sqrt{3}}\,f_{L}^{2 2} - 2\,f_{L,\,y_1}^{0 0} %\nonumber\\
%&&\hspace{4.1cm}- 
  - {\sqrt{2}}\,f_{L,\,y_1}^{2 0} + {\sqrt{3}}\,f_{L,\,y_1}^{2 2}\Big) \,.
\label{t66}
\eeqa

\begin{figure}
\centerline{%
\epsfxsize=9.0cm
\epsffile{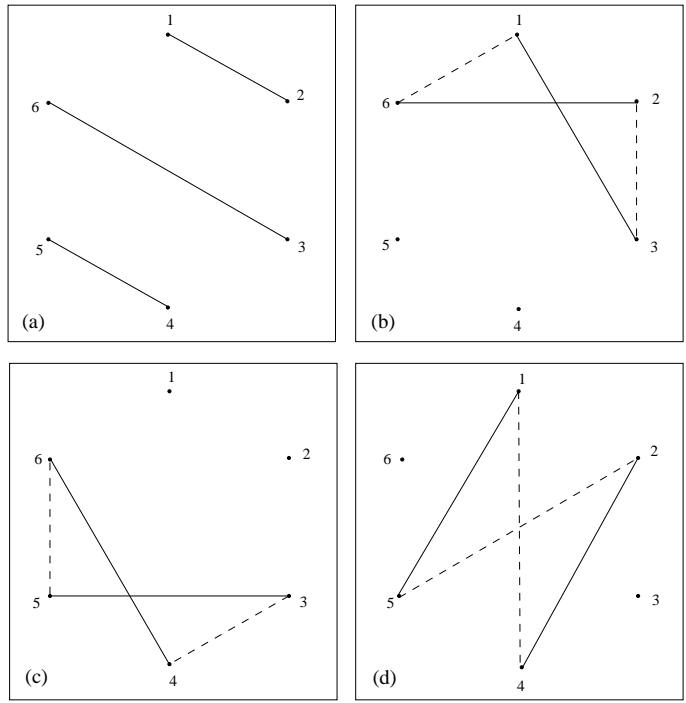}
}
\vspace*{1cm}
\caption{As Fig.\ \protect{\ref{figLhel}} for the transformed helicity 
basis with nomenclature in Table \protect{\ref{tabgroupLhel4}}. 
\label{figLhel4}
}
\end{figure}

\begin{figure}
\centerline{%
\epsfxsize=13.0cm
\epsffile{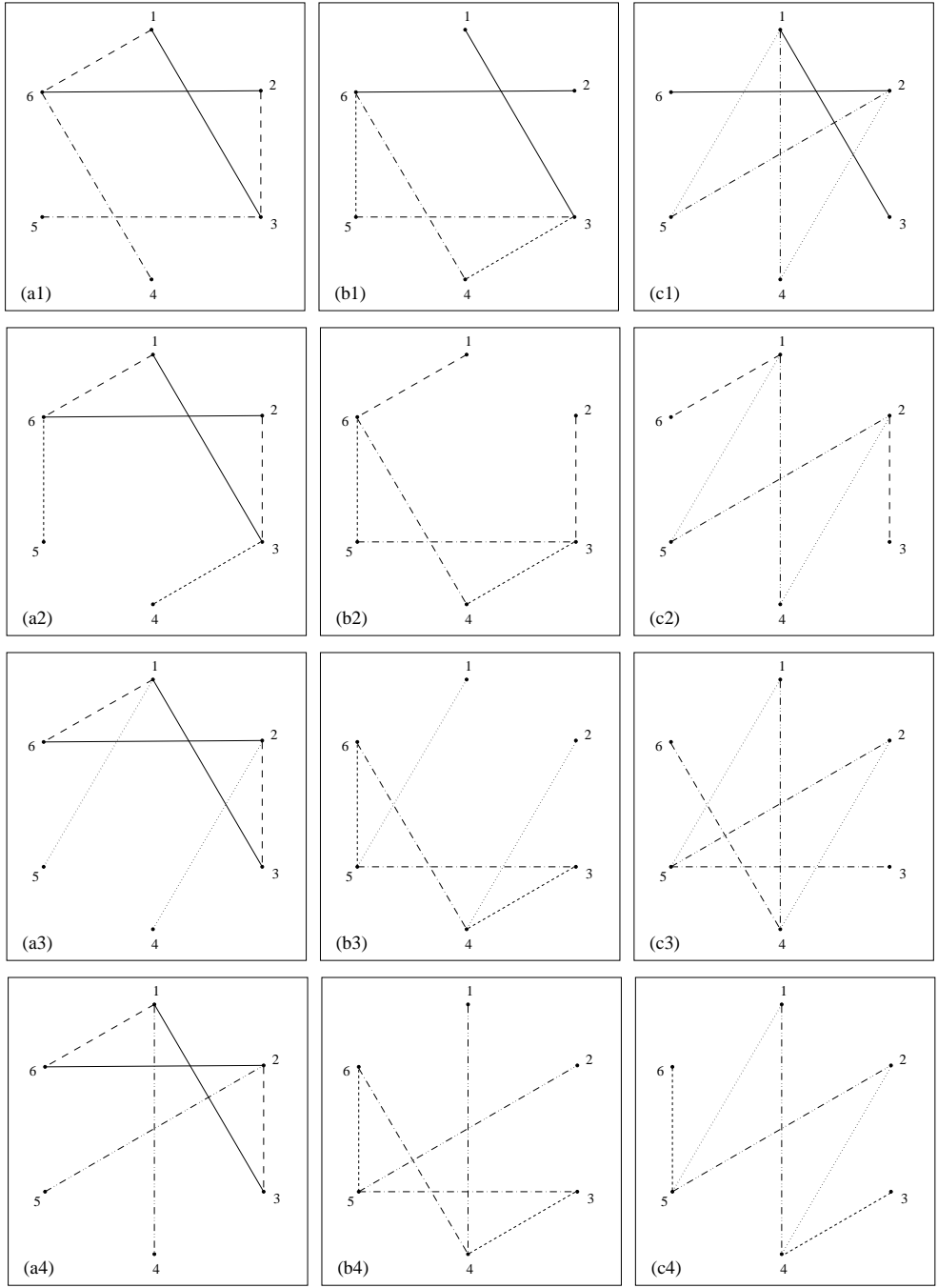}
}
\vspace*{1cm}
\caption{Connected diagrams with three groups of longitudinal observables 
describing six different interference terms and containing closed loops 
with four points allowing a complete determination of five 
\protect{$t$}-matrix elements as function of the remaining one for the 
transformed helicity basis where $f_L^{IM}(X)$ is represented by $X^{IM}$:
          solid: $1^{21}$, $1^{11}$, $y_1^{11}$, $y_1^{21}$, 
    long dashed: $xz^{11}$, $xz^{21}$, $zz^{11}$, $zz^{21}$, 
    dash-dotted: $x_1^{10}$, $x_1^{22}$, $z_1^{10}$, $z_1^{22}$, 
   short dashed: $x_2^{10}$, $x_2^{22}$, $z_2^{10}$, $z_2^{22}$, 
         dotted: $x_1^{11}$, $x_1^{21}$, $z_1^{11}$, $z_1^{21}$, 
dash-double-dot: $x_2^{11}$, $x_2^{21}$, $z_2^{11}$, $z_2^{21}$.
\label{4loop}
}
\end{figure}

For obvious reasons we will choose for this case only groups of two interference 
terms which are represented by four observables, i.e., the six groups represented in 
the panels (b) through (d) in Fig.\ \ref{figLhel4}. Avoiding as mentioned above 
closed loops of three, one finds three closed loops containing four points, already
shown in Fig.\ \ref{figLhel4}. Each of these loops can be combined with one of the 
remaining four groups in order to find a minimal pattern for the determination of 
all matrix elements by one. Thus one finds altogether twelve different 
combinations of three groups containing one four-loop shown diagrammatically in 
Fig.\ \ref{4loop}. Let us consider explicitly the case of panel (a1). 
It involves the group
$\{T_{3,1},\,T_{6,2}\}$, $\{T_{6,1},\,T_{3,2}\}$, and $\{T_{5,3},\,T_{6,4}\}$. 
We will choose $t_1$ as real and positive and express all other matrix elements 
in terms of $t_1$ as follows (note $T_{i,j}=T_{j,i}^*$)
\beqa
t_3&=&\frac{1}{t_1}\,T_{1,3}\,,\qquad
t_6=\frac{1}{t_1}\,T_{1,6}\,,\label{t36}\\
t_2&=&t_1\,\frac{T_{6,2}}{T_{6,1}}\,,\qquad
t_4=t_1\,\frac{T_{6,4}}{T_{6,1}}\,,\\
t_5&=&t_1\,\frac{T_{3,5}}{T_{3,1}}\,.\label{t5}
\eeqa
Because of the closed four-loop one has from (\ref{evenpoints1}) the condition
\beq
T_{1,6}T_{2,3}=T_{2,6}T_{1,3}\,,
\eeq
which can be used to eliminate two of the eight observables of the groups 
$\{T_{6,1},\,T_{3,2}\}$ and $\{T_{3,1},\,T_{6,2}\}$. Formally, there is 
no preference as to which observables one should eliminate. A closer 
inspection of the explicit expressions of the $T_{i,j}$ of each one of the 
three groups (see Eq.\ (\ref{t66})) reveals that the general structure 
for the two interference terms belonging to one group is of the form
\beq
T_{i,j}= a_1+a_2+i(b_1+b_2)\,,\qquad T_{i',j'}= a_1-a_2+i(b_1-b_2)\,,\label{generalform}
\eeq
where $a_i$ and $b_i$ are observables up to some constants, and thus are real quantities.
Therefore, the above relation can be written in the general form separating real and 
imaginary parts and denoting the observables of one group by $\{a_i,b_i\}$ and 
those of the other group by $\{\bar a_i,\bar b_i\}$
\beqa
a_1^2-a_2^2 -b_1^2+b_2^2&=&\bar a_1^2-\bar a_2^2 -\bar b_1^2+\bar b_2^2\,,\\
a_1 b_1 - a_2 b_2 &=& \bar a_1 \bar b_1 - \bar a_2 \bar b_2\,,
\eeqa
by which two observables can be eliminated. For example, $a_1$ and $b_1$ can be 
eliminated by
\beqa
b_1&=& \pm \sqrt{\frac{1}{2}(|\alpha+ 2i\beta|-\alpha)}\,,\\
a_1 &=& \frac{\beta}{b_1}\,,
\eeqa
where 
\beqa
\alpha&=& \bar a_1^2-\bar a_2^2 -\bar b_1^2+\bar b_2^2 + a_2^2- b_2^2\,,\\
\beta&=& \bar a_1 \bar b_1 - \bar a_2 \bar b_2 + a_2 b_2\,.
\eeqa
Thus all five complex matrix elements 
in (\ref{t36}) through (\ref{t5}) are determined by ten observables and by $t_1$. 
The latter one can be determined finally, for example, from 
\beq
f_L = (t_1)^2\,\Big(1 + \Big|\frac{T_{6,2}}{T_{6,1}}\Big|^2 
      + \Big|\frac{T_{6,4}}{T_{6,1}}\Big|^2 
      + \Big|\frac{T_{3,5}}{T_{3,1}}\Big|^2\Big) 
      + \frac{1}{|t_1|^2}\,\Big(|T_{1,3}|^2 + |T_{1,6}|^2  \Big)\,,
\eeq
which in general will provide two solutions. 

\begin{figure}
\centerline{%
\epsfxsize=9.0cm
\epsffile{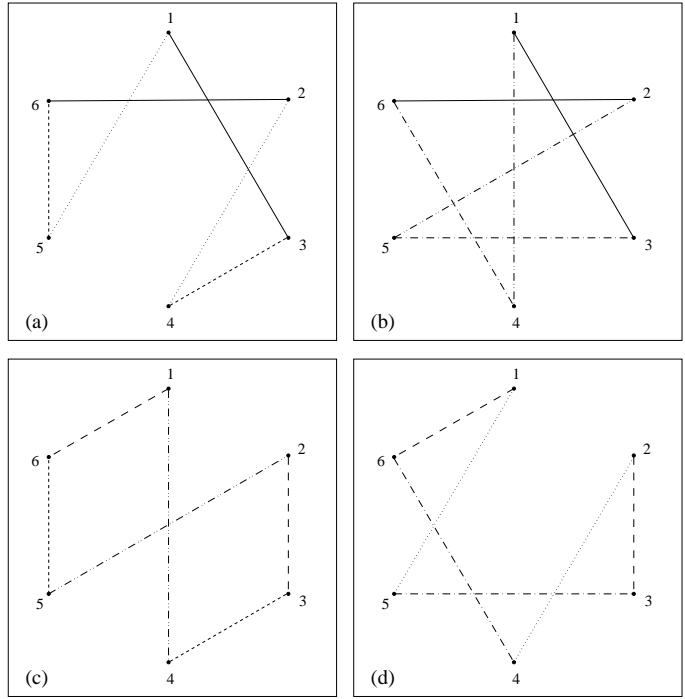}
}
\vspace*{1cm}
\caption{As Fig.\ \protect{\ref{4loop}} containing closed loops with six points: 
          solid: $1^{21}$, $1^{11}$, $y_1^{11}$, $y_1^{21}$, 
    long dashed: $xz^{11}$, $xz^{21}$, $zz^{11}$, $zz^{21}$, 
    dash-dotted: $x_1^{10}$, $x_1^{22}$, $z_1^{10}$, $z_1^{22}$, 
   short dashed: $x_2^{10}$, $x_2^{22}$, $z_2^{10}$, $z_2^{22}$, 
         dotted: $x_1^{11}$, $x_1^{21}$, $z_1^{11}$, $z_1^{21}$, 
dash-double-dot: $x_2^{11}$, $x_2^{21}$, $z_2^{11}$. $z_2^{21}$, 
\label{6loop}
}
\end{figure}

Another possibility is to choose closed loops consisting of all six points. 
These are represented in  Fig.\ \ref{6loop}. As an example, we will discuss 
the case of panel (a) containing only groups which are governed 
by single and double polarization observables alone. The corresponding interference 
terms of these groups are given by $\{T_{3,1},\,T_{6,2}\}$, $\{T_{5,1},\,T_{4,2}\}$, 
and $\{T_{4,3},\,T_{6,5}\}$ listed above. Again we will choose $t_1$ as real and positive 
and express all other matrix elements in terms of $t_1$ 
\beqa
t_3&=&\frac{1}{t_1}\,T_{1,3}\,,\qquad
t_5=\frac{1}{t_1}\,T_{1,5}\,,\label{t35}\\
t_4&=&t_1\,\frac{T_{3,4}}{T_{3,1}}\,,\qquad
t_6=t_1\,\frac{T_{5,6}}{T_{5,1}}\,,\\
t_2&=&\frac{1}{t_1}\,\frac{T_{1,3}T_{4,2}}{T_{4,3}}\,.\label{t2}
\eeqa
Furthermore, the closed loop through all six points leads to the following relation 
between observables
\beq
T_{3,4}\,T_{5,1}\,T_{2,6}=T_{5,6}\,T_{3,1}\,T_{2,4}\,.\label{Trelation}
\eeq
It can be used to eliminate two of the twelve observables involved belonging to one group. 
For example, one may write 
\beq
T_{3,4}=T_{5,6}\,T\quad \mbox{with}\quad T=\frac{T_{3,1}\,T_{2,4}}{T_{5,1}\,T_{2,6}}\,,
\label{Trelation56}
\eeq
where $T_{3,4}$ and $T_{5,6}$ belong to one group of observables which do not appear in $T$.

Also here, there is no preference as to which observables one should eliminate, and 
again the two interference terms belonging to one group and appearing in 
(\ref{Trelation56}) are of the form
\beq
T_{i,j}= a_1+a_2+i(b_1+b_2)\,,\qquad T_{i',j'}= a_1-a_2-i(b_1-b_2)\,.
\eeq
Thus the above relation leads to 
\beq
a_1+a_2+i(b_1+b_2)=(a_1-a_2-i(b_1-b_2))\,T\,,\label{Trelation1}
\eeq
where $T$ is a ratio of the interference terms of the other two groups depending on 
the choice of the two interference terms $T_{i,j}$ and $T_{i',j'}$ of one group 
(see for example (\ref{Trelation56})). It is important to note 
that $T$ does not depend on the observables of $a_i$ and $b_i$. Then 
taking the real and imaginary parts of (\ref{Trelation1}), one obtains a system of linear 
equations between the $a_i$ and $b_i$
\beqa
a_1+a_2 &=& \Re e (T)\,(a_1-a_2)+ \Im m (T)\,(b_1-b_2)\,,\\
b_1+b_2 &=& \Im m (T)\,(a_1-a_2)- \Re e (T)\,(b_1-b_2)\,,
\eeqa
which can be used to eliminate two of the four observables. For example, if one wants 
to eliminate $a_i$, one easily finds
\beqa
a_1-a_2 &=& \frac{1}{\Im m (T)}(b_1+b_2 + \Re e (T)\,(b_1-b_2))\,,\\
a_1+a_2 &=& \frac{1}{\Im m (T)}(\Re e (T)\,(b_1+b_2)+ |T|^2 (b_1-b_2))\,,
\eeqa
provided $\Im m (T)\neq 0$.
The corresponding solution for $b_i$ is obtained from this by exchanging $a_i 
\leftrightarrow b_i$ and changing the sign of $\Im m (T)$. If one wants to eliminate 
$a_1$ and $b_1$ one obtains
\beqa
a_1 &=& \frac{1}{|T|^2-1}\Big(|1+T|^2\,a_2 +2\Im m (T)\,b_2\Big)\,,\\
b_1 &=& \frac{1}{|T|^2-1}\Big(2\Im m (T)\,a_2 +|1+T|^2\,b_2\Big)\,
\eeqa
provided $|T|^2\neq 1$. Thus all five complex matrix elements in (\ref{t35})
through (\ref{t2}) are determined by ten observables and by $t_1$, while 
the latter one can be determined, for example, as above from 
\beq
f_L = (t_1)^2\,\Big(1 + \Big|\frac{T_{3,4}}{T_{3,1}}\Big|^2 
      + \Big|\frac{T_{5,6}}{T_{5,1}}\Big|^2\Big) 
      + \frac{1}{|t_1|^2}\,\Big(|T_{1,3}|^2 + |T_{1,5}|^2  
      + \Big|\frac{T_{1,3}T_{4,2}}{T_{4,3}}\Big|^2 \Big)\,,
\eeq
with again two solutions in general. 

\begin{figure}
\centerline{%
\epsfxsize=9.0cm
\epsffile{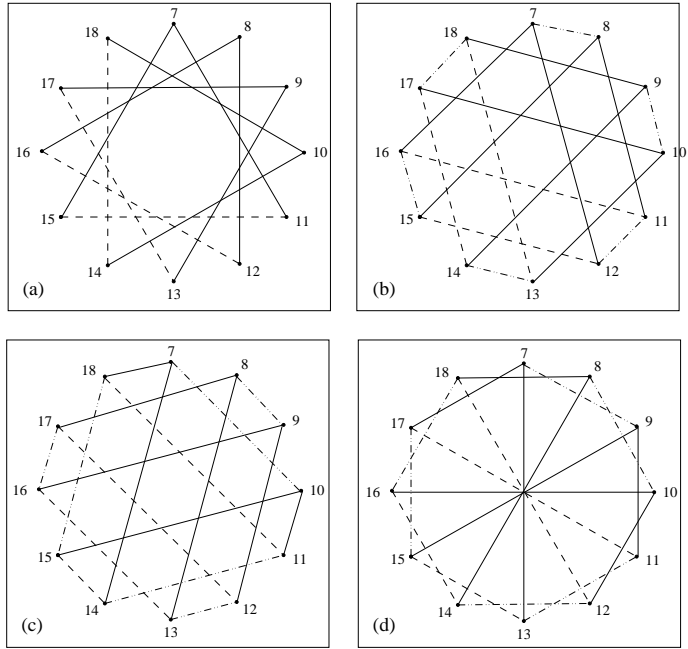}
}
\vspace*{1cm}
\caption{Diagrammatic representation of groups of transverse observables 
determining the interference terms of $t$-matrix elements for the  helicity 
and the transformed helicity basis. 
The nomenclature for the groups and the corresponding observables are 
listed in Table \protect{\ref{tabgroupThel}} for the helicity basis and 
in Table \protect{\ref{tabgroupThel4}} for the transformed helicity basis.
\label{figThel}
}
\end{figure}

Next we will consider the more involved transverse matrix elements. In this 
case the rotation of the spin quantization axis for the initial and/or final state 
does not lead to a simpler grouping of observables than shown in Fig.\ \ref{figThel}. 
Thus we will use the helicity basis in the following discussion. But the results 
apply as well to the transformed helicity basis discussed before for the longitudinal 
case, although for different groups of observables. A closer inspection of the 
various types of groups shows that in order to obtain a diagrammatic pattern in 
which each point is connected to the others, not necessarily directly, one needs 
a proper combination of all three types of groups as 
distinguished in Fig.\ \ref{figThel} by the different line types. Only the group 
represented by dashed lines in panel (a) of Fig.\ \ref{figThel} has to be excluded, 
because it always leads with the solid-line groups to triangular loops. Thus the 
possible combinations of all four solid-line groups (panels (a) through (d)) 
with all three dashed-line and three dash-dot-dot-line groups (panels (b) through 
(d)) amount to 36 different patterns. 

\begin{figure}
\centerline{%
\epsfxsize=13.0cm
\epsffile{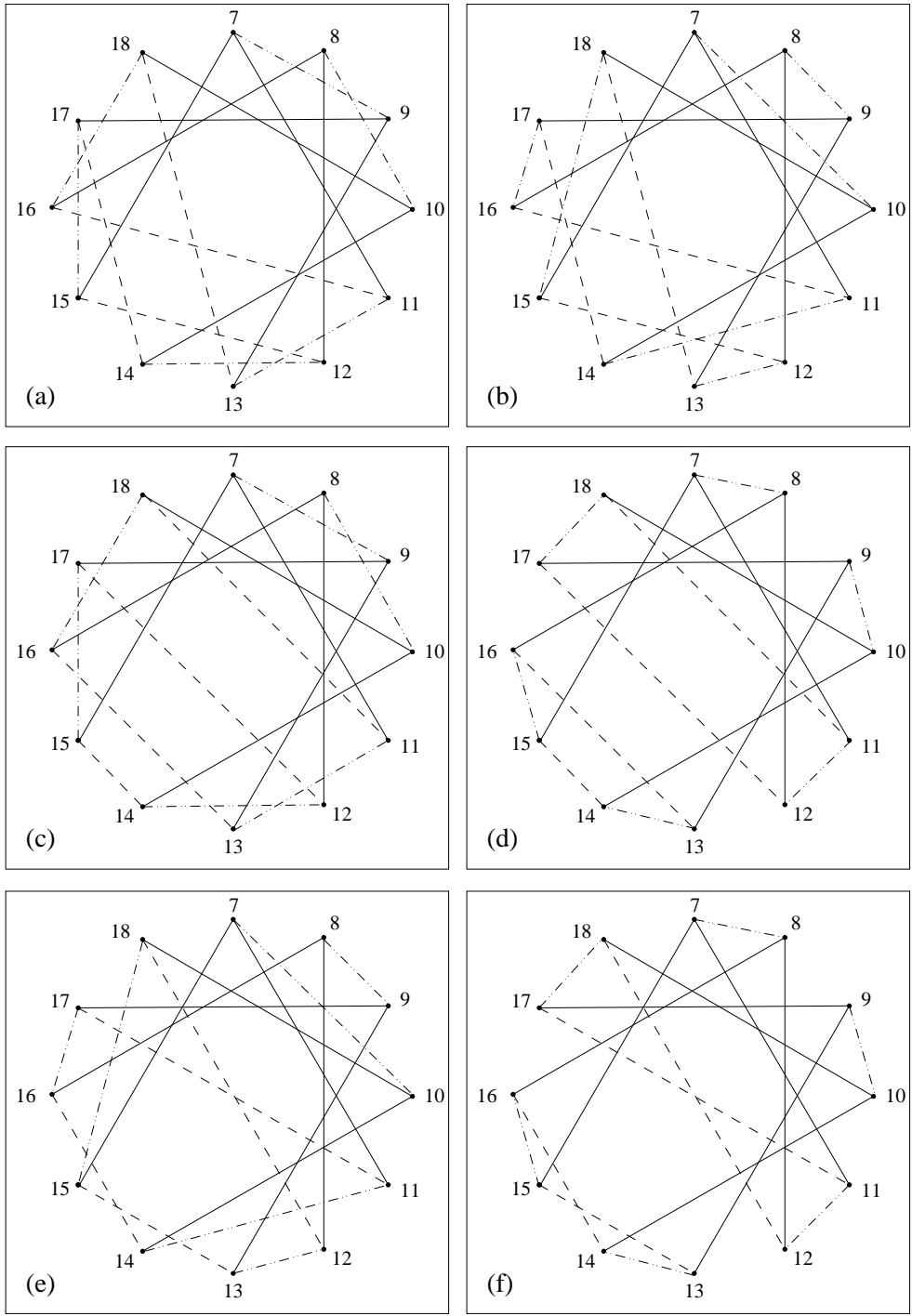}
}
\vspace*{1cm}
\caption{Diagrammatic representation of three groups of transverse observables 
leading to a connected diagram for all interference terms 
allowing a complete determination of eleven \protect{$t$}-matrix elements as 
function of the remaining one for the helicity or the transformed helicity 
basis. The corresponding observables are explained in the text. 
\label{connpat1}
}
\end{figure}

However those combinations where dashed lines run parallel to dash-dot-dot lines 
also have to be excluded, because they result in patterns with two disconnected 
groups of points. Three such combinations appear in the panels (b) through 
(d) of Fig.\ \ref{figThel}. Therefore, 12 of the 36 patterns have to be excluded, 
leaving 24 connected patterns, which can be used for the determination of 
all transverse $t$-matrix elements in terms of one. We show in Fig.\ \ref{connpat1}
for the solid-line group of panel (a) in Fig.\ \ref{figThel} the resulting six 
connected diagrams of which we will choose the one in panel (a) for a closer analysis. 
All 24 connected patterns are governed by 36 observables (16 for the solid-line groups, 12 
for the dash-dot-dot ones, and 8 for the dashed ones, see Table \ref{tabgroupThel}). 
As we will soon see, one finds six independent four-point-loops and one independent 
six-point-loop, through which one can eliminate 14 observables, thus resulting in 
22 observables, which is just the minimal number required to express all matrix 
elements in terms of one. 

Now we will turn to a detailed discussion of the pattern in panel (a) of 
Fig.\ \ref{connpat1}. The diagram consists of the following groups of observables 
listed in Table \ref{tabgroupThel}: solid lines of panel (a), dashed lines of panel (b), 
and dash-dot-dot lines of panel (d). The explicit bilinear expressions for the 
solid-line group are 
\beqa
T_{11,7} & = & t^*_{{1\over 2} {1\over 2} 1 1} t_{
  {1\over 2} {1\over 2} 1 0} \nonumber\\
 & = & \frac{1}{16\sqrt{6}}\Big(-f_{T}^{2 1} 
  + f_{T}^{\prime\, 1 1} - 
   f_{T,\,zz}^{2 1} + f_{T,\,zz}^{\prime\, 1 1} - f_{T,\,z_1}^{1 1} + 
   f_{T,\,z_1}^{\prime\, 2 1} - f_{T,\,z_2}^{1 1} + 
   f_{T,\,z_2}^{\prime\, 2 1}\nonumber\\
& & +
  i\,\left( -f_{T}^{1 1} - f_{T}^{\prime\, 2 1} - f_{T,\,zz}^{1 1} - 
     f_{T,\,zz}^{\prime\, 2 1} - f_{T,\,z_1}^{2 1} - 
     f_{T,\,z_1}^{\prime\, 1 1} - f_{T,\,z_2}^{2 1} - 
     f_{T,\,z_2}^{\prime\, 1 1} \right) \Big)\,,\nonumber\\
T_{15,7} & = & t^*_{{1\over 2} {1\over 2} 1 -1} t_{
  {1\over 2} {1\over 2} 1 0}  \nonumber\\
 & = & \frac{1}{16\sqrt{6}}\Big(f_{T}^{2 1} 
  + f_{T}^{\prime\, 1 1} + 
   f_{T,\,zz}^{2 1} + f_{T,\,zz}^{\prime\, 1 1} - f_{T,\,z_1}^{1 1} - 
   f_{T,\,z_1}^{\prime\, 2 1} - f_{T,\,z_2}^{1 1} - 
   f_{T,\,z_2}^{\prime\, 2 1} \nonumber\\
& & +
  i\,\left( f_{T}^{1 1} - f_{T}^{\prime\, 2 1} + f_{T,\,zz}^{1 1} - 
     f_{T,\,zz}^{\prime\, 2 1} - f_{T,\,z_1}^{2 1} + 
     f_{T,\,z_1}^{\prime\, 1 1} - f_{T,\,z_2}^{2 1} + 
     f_{T,\,z_2}^{\prime\, 1 1} \right) \Big)\,,\nonumber\\
T_{12,8} & = & t^*_{{1\over 2} -{1\over 2} 1 1} t_{
  {1\over 2} -{1\over 2} 1 0}  \nonumber\\
 & = & \frac{1}{16\sqrt{6}}\Big(
  -f_{T}^{2 1} + f_{T}^{\prime\, 1 1} + f_{T,\,zz}^{2 1} - 
   f_{T,\,zz}^{\prime\, 1 1} - f_{T,\,z_1}^{1 1} + 
   f_{T,\,z_1}^{\prime\, 2 1} + f_{T,\,z_2}^{1 1} - 
   f_{T,\,z_2}^{\prime\, 2 1} \nonumber\\
& & +
  i\,\left( -f_{T}^{1 1} - f_{T}^{\prime\, 2 1} + f_{T,\,zz}^{1 1} + 
     f_{T,\,zz}^{\prime\, 2 1} - f_{T,\,z_1}^{2 1} - 
     f_{T,\,z_1}^{\prime\, 1 1} + f_{T,\,z_2}^{2 1} + 
     f_{T,\,z_2}^{\prime\, 1 1} \right) \Big)\,,\nonumber\\
T_{16,8} & = & t^*_{{1\over 2} -{1\over 2} 1 -1} t_{
  {1\over 2} -{1\over 2} 1 0}  \nonumber\\
 & = & \frac{1}{16\sqrt{6}}\Big(
  f_{T}^{2 1} + f_{T}^{\prime\, 1 1} - f_{T,\,zz}^{2 1} - 
   f_{T,\,zz}^{\prime\, 1 1} - f_{T,\,z_1}^{1 1} - 
   f_{T,\,z_1}^{\prime\, 2 1} + f_{T,\,z_2}^{1 1} + 
   f_{T,\,z_2}^{\prime\, 2 1} \nonumber\\
& & +
  i\,\left( f_{T}^{1 1} - f_{T}^{\prime\, 2 1} - f_{T,\,zz}^{1 1} + 
     f_{T,\,zz}^{\prime\, 2 1} - f_{T,\,z_1}^{2 1} + 
     f_{T,\,z_1}^{\prime\, 1 1} + f_{T,\,z_2}^{2 1} - 
     f_{T,\,z_2}^{\prime\, 1 1} \right) \Big)\,,\nonumber\\
T_{13,9} & = & t^*_{-{1\over 2} {1\over 2} 1 1} t_{
  -{1\over 2} {1\over 2} 1 0}  \nonumber\\
 & = & \frac{1}{16\sqrt{6}}\Big(
  -f_{T}^{2 1} + f_{T}^{\prime\, 1 1} + f_{T,\,zz}^{2 1} - 
   f_{T,\,zz}^{\prime\, 1 1} + f_{T,\,z_1}^{1 1} - 
   f_{T,\,z_1}^{\prime\, 2 1} - f_{T,\,z_2}^{1 1} + 
   f_{T,\,z_2}^{\prime\, 2 1} \nonumber\\
& & +
  i\,\left( -f_{T}^{1 1} - f_{T}^{\prime\, 2 1} + f_{T,\,zz}^{1 1} + 
     f_{T,\,zz}^{\prime\, 2 1} + f_{T,\,z_1}^{2 1} + 
     f_{T,\,z_1}^{\prime\, 1 1} - f_{T,\,z_2}^{2 1} - 
     f_{T,\,z_2}^{\prime\, 1 1} \right) \Big)\,,\nonumber\\
T_{17,9} & = & t^*_{-{1\over 2} {1\over 2} 1 -1} t_{
  -{1\over 2} {1\over 2} 1 0}  \nonumber\\
 & = & \frac{1}{16\sqrt{6}}\Big(
  f_{T}^{2 1} + f_{T}^{\prime\, 1 1} - f_{T,\,zz}^{2 1} - 
   f_{T,\,zz}^{\prime\, 1 1} + f_{T,\,z_1}^{1 1} + 
   f_{T,\,z_1}^{\prime\, 2 1} - f_{T,\,z_2}^{1 1} - 
   f_{T,\,z_2}^{\prime\, 2 1} \nonumber\\
& & +
  i\,\left( f_{T}^{1 1} - f_{T}^{\prime\, 2 1} - f_{T,\,zz}^{1 1} + 
     f_{T,\,zz}^{\prime\, 2 1} + f_{T,\,z_1}^{2 1} - 
     f_{T,\,z_1}^{\prime\, 1 1} - f_{T,\,z_2}^{2 1} + 
     f_{T,\,z_2}^{\prime\, 1 1} \right) \Big)\,,\nonumber\\
T_{14,10} & = & t^*_{-{1\over 2} -{1\over 2} 1 1} t_{
  -{1\over 2} -{1\over 2} 1 0}  \nonumber\\
 & = & \frac{1}{16\sqrt{6}}\Big(
  -f_{T}^{2 1} + f_{T}^{\prime\, 1 1} - f_{T,\,zz}^{2 1} + 
   f_{T,\,zz}^{\prime\, 1 1} + f_{T,\,z_1}^{1 1} - 
   f_{T,\,z_1}^{\prime\, 2 1} + f_{T,\,z_2}^{1 1} - 
   f_{T,\,z_2}^{\prime\, 2 1} \nonumber\\
& & +
  i\,\left( -f_{T}^{1 1} - f_{T}^{\prime\, 2 1} - f_{T,\,zz}^{1 1} - 
     f_{T,\,zz}^{\prime\, 2 1} + f_{T,\,z_1}^{2 1} + 
     f_{T,\,z_1}^{\prime\, 1 1} + f_{T,\,z_2}^{2 1} + 
     f_{T,\,z_2}^{\prime\, 1 1} \right) \Big)\,,\nonumber\\
T_{18,10} & = & t^*_{-{1\over 2} -{1\over 2} 1 -1
  } t_{-{1\over 2} -{1\over 2} 1 0}  \nonumber\\
 & = & \frac{1}{16\sqrt{6}}\Big(
  f_{T}^{2 1} + f_{T}^{\prime\, 1 1} + f_{T,\,zz}^{2 1} + 
   f_{T,\,zz}^{\prime\, 1 1} + f_{T,\,z_1}^{1 1} + 
   f_{T,\,z_1}^{\prime\, 2 1} + f_{T,\,z_2}^{1 1} + 
   f_{T,\,z_2}^{\prime\, 2 1} \nonumber\\
& & +
  i\,\left( f_{T}^{1 1} - f_{T}^{\prime\, 2 1} + f_{T,\,zz}^{1 1} - 
     f_{T,\,zz}^{\prime\, 2 1} + f_{T,\,z_1}^{2 1} - 
     f_{T,\,z_1}^{\prime\, 1 1} + f_{T,\,z_2}^{2 1} - 
     f_{T,\,z_2}^{\prime\, 1 1} \right) \Big)\,,\nonumber 
\eeqa  
for the dashed-line group
\beqa
T_{16,11} & = & t^*_{{1\over 2} -{1\over 2} 1 -1} t_{
  {1\over 2} {1\over 2} 1 1} \nonumber\\
 & = & \frac{1}{16\sqrt{3}}\Big(-f_{T,\,x_2}^{\prime\, 2 2} - 
   f_{T,\,y_2}^{\prime\, 2 2} + f_{T,\,zx}^{2 2} - f_{T,\,zy}^{2 2}  +
  i\,\left( -f_{T,\,x_2}^{2 2} - f_{T,\,y_2}^{2 2} - 
     f_{T,\,zx}^{\prime\, 2 2} + f_{T,\,zy}^{\prime\, 2 2} \right) \Big)\,,
\nonumber\\
T_{15,12} & = & t^*_{{1\over 2} {1\over 2} 1 -1} t_{
  {1\over 2} -{1\over 2} 1 1} \nonumber\\
 & = & \frac{1}{16\sqrt{3}}\Big(
  -f_{T,\,x_2}^{\prime\, 2 2} + f_{T,\,y_2}^{\prime\, 2 2} + 
   f_{T,\,zx}^{2 2} + f_{T,\,zy}^{2 2}  +
  i\,\left( -f_{T,\,x_2}^{2 2} + f_{T,\,y_2}^{2 2} - 
     f_{T,\,zx}^{\prime\, 2 2} - f_{T,\,zy}^{\prime\, 2 2} \right) \Big)\,,
\nonumber\\
T_{18,13} & = & t^*_{-{1\over 2} -{1\over 2} 1 -1
  } t_{-{1\over 2} {1\over 2} 1 1} \nonumber\\
 & = & \frac{1}{16\sqrt{3}}\Big(
  -f_{T,\,x_2}^{\prime\, 2 2} - f_{T,\,y_2}^{\prime\, 2 2} - 
   f_{T,\,zx}^{2 2} + f_{T,\,zy}^{2 2}  +
  i\,\left( -f_{T,\,x_2}^{2 2} - f_{T,\,y_2}^{2 2} + 
     f_{T,\,zx}^{\prime\, 2 2} - f_{T,\,zy}^{\prime\, 2 2} \right) \Big)\,,
\nonumber\\
T_{17,14} & = & t^*_{-{1\over 2} {1\over 2} 1 -1} t_{
  -{1\over 2} -{1\over 2} 1 1} \nonumber\\
 & = & \frac{1}{16\sqrt{3}}\Big(
  -f_{T,\,x_2}^{\prime\, 2 2} + f_{T,\,y_2}^{\prime\, 2 2} - 
   f_{T,\,zx}^{2 2} - f_{T,\,zy}^{2 2}  +
  i\,\left( -f_{T,\,x_2}^{2 2} + f_{T,\,y_2}^{2 2} + 
     f_{T,\,zx}^{\prime\, 2 2} + f_{T,\,zy}^{\prime\, 2 2} \right) \Big)\,,
\nonumber
\eeqa  
and for the dash-dot-dot-line group
\beqa  
T_{9,7} & = & t^*_{-{1\over 2} {1\over 2} 1 0} t_{{1\over 2} 
  {1\over 2} 1 0} \nonumber\\
 & = & \frac{1}{24}\Big(f_{T,\,xz}^{0 0} - {\sqrt{2}}\,f_{T,\,xz}^{2 0} - 
   f_{T,\,x_1}^{\prime\, 0 0} + {\sqrt{2}}\,f_{T,\,x_1}^{\prime\, 2 0} 
%\nonumber\\
%& & +
 + i\,\left( f_{T,\,yz}^{\prime\, 0 0} - 
     {\sqrt{2}}\,f_{T,\,yz}^{\prime\, 2 0} - f_{T,\,y_1}^{0 0} + 
     {\sqrt{2}}\,f_{T,\,y_1}^{2 0} \right) \Big)\,,\nonumber\\
T_{10,8} & = & t^*_{-{1\over 2} -{1\over 2} 1 0} t_{{1\over 2} 
  -{1\over 2} 1 0} \nonumber\\
 & = & \frac{1}{24}\Big(-f_{T,\,xz}^{0 0} + {\sqrt{2}}\,f_{T,\,xz}^{2 0} - 
   f_{T,\,x_1}^{\prime\, 0 0} + {\sqrt{2}}\,f_{T,\,x_1}^{\prime\, 2 0} 
%\nonumber\\
%& &  +
 + i\,\left( -f_{T,\,yz}^{\prime\, 0 0} + 
     {\sqrt{2}}\,f_{T,\,yz}^{\prime\, 2 0} - f_{T,\,y_1}^{0 0} + 
     {\sqrt{2}}\,f_{T,\,y_1}^{2 0} \right) \Big)\,,\nonumber\\
T_{13,11} & = &t^*_{-{1\over 2} {1\over 2} 1 1} t_{
  {1\over 2} {1\over 2} 1 1} \nonumber\\
 & = & \frac{1}{24\sqrt{2}}\Big({\sqrt{2}}\,f_{T,\,xz}^{0 0} + 
   f_{T,\,xz}^{2 0} - {\sqrt{3}}\,f_{T,\,xz}^{\prime\, 1 0} + 
   {\sqrt{3}}\,f_{T,\,x_1}^{1 0} - {\sqrt{2}}\,f_{T,\,x_1}^{\prime\, 0 0} - 
   f_{T,\,x_1}^{\prime\, 2 0} \nonumber\\
& & +
  i\,\left( -{\sqrt{3}}\,f_{T,\,yz}^{1 0}  + 
     {\sqrt{2}}\,f_{T,\,yz}^{\prime\, 0 0} + f_{T,\,yz}^{\prime\, 2 0} - 
     {\sqrt{2}}\,f_{T,\,y_1}^{0 0} - f_{T,\,y_1}^{2 0} + 
     {\sqrt{3}}\,f_{T,\,y_1}^{\prime\, 1 0} \right) \Big)\,,\nonumber\\
T_{14,12} & = &t^*_{-{1\over 2} -{1\over 2} 1 1} t_{
  {1\over 2} -{1\over 2} 1 1} \nonumber\\
 & = & \frac{1}{24\sqrt{2}}\Big(
  - {\sqrt{2}}\,f_{T,\,xz}^{0 0}  - f_{T,\,xz}^{2 0} + 
   {\sqrt{3}}\,f_{T,\,xz}^{\prime\, 1 0} + {\sqrt{3}}\,f_{T,\,x_1}^{1 0} - 
   {\sqrt{2}}\,f_{T,\,x_1}^{\prime\, 0 0} - f_{T,\,x_1}^{\prime\, 2 0} \nonumber\\
& & +
  i\,\left( {\sqrt{3}}\,f_{T,\,yz}^{1 0} - 
     {\sqrt{2}}\,f_{T,\,yz}^{\prime\, 0 0} - f_{T,\,yz}^{\prime\, 2 0} - 
     {\sqrt{2}}\,f_{T,\,y_1}^{0 0} - f_{T,\,y_1}^{2 0} + 
     {\sqrt{3}}\,f_{T,\,y_1}^{\prime\, 1 0} \right) \Big)\,,\nonumber\\
T_{17,15} & = &t^*_{-{1\over 2} {1\over 2} 1 -1} t_{
  {1\over 2} {1\over 2} 1 -1} \nonumber\\
 & = & \frac{1}{24\sqrt{2}}\Big(
  {\sqrt{2}}\,f_{T,\,xz}^{0 0} + f_{T,\,xz}^{2 0} + 
   {\sqrt{3}}\,f_{T,\,xz}^{\prime\, 1 0} - {\sqrt{3}}\,f_{T,\,x_1}^{1 0} - 
   {\sqrt{2}}\,f_{T,\,x_1}^{\prime\, 0 0} - f_{T,\,x_1}^{\prime\, 2 0} \nonumber\\
& & +
  i\,\left( {\sqrt{3}}\,f_{T,\,yz}^{1 0} + 
     {\sqrt{2}}\,f_{T,\,yz}^{\prime\, 0 0} + f_{T,\,yz}^{\prime\, 2 0} - 
     {\sqrt{2}}\,f_{T,\,y_1}^{0 0} - f_{T,\,y_1}^{2 0} - 
     {\sqrt{3}}\,f_{T,\,y_1}^{\prime\, 1 0} \right) \Big)\,,\nonumber\\
T_{18,16} & = &t^*_{-{1\over 2} -{1\over 2} 1 -1
  } t_{{1\over 2} -{1\over 2} 1 -1} \nonumber\\
 & = & \frac{1}{24\sqrt{2}}\Big(
  -{\sqrt{2}}\,f_{T,\,xz}^{0 0} - f_{T,\,xz}^{2 0} - 
   {\sqrt{3}}\,f_{T,\,xz}^{\prime\, 1 0} - {\sqrt{3}}\,f_{T,\,x_1}^{1 0} - 
   {\sqrt{2}}\,f_{T,\,x_1}^{\prime\, 0 0} - f_{T,\,x_1}^{\prime\, 2 0} \nonumber\\
& & +
  i\,\left( -{\sqrt{3}}\,f_{T,\,yz}^{1 0}  - 
     {\sqrt{2}}\,f_{T,\,yz}^{\prime\, 0 0} - f_{T,\,yz}^{\prime\, 2 0} - 
     {\sqrt{2}}\,f_{T,\,y_1}^{0 0} - f_{T,\,y_1}^{2 0} - 
     {\sqrt{3}}\,f_{T,\,y_1}^{\prime\, 1 0} \right) \Big)\,.\nonumber
\eeqa  

One finds six independent 4-point loops, namely (7-9-17-15), (7-9-13-11), 
(8-10-14-12), (8-10-18-16), (11-13-18-16), and (12-14-17-15). They all involve two 
dash-dot-dot-lines and lead to the following relations according to 
(\ref{evenpoints1})
\beqa
T_{15,17}T_{9,7} &=& T_{9,17}T_{15,7}\,, \quad \mbox{for (7-9-17-15)}
\,,\label{4L1}\\
T_{11,13}T_{9,7} &=& T_{9,13}T_{11,7}\,, \quad \mbox{for (7-9-13-11)}
\,,\label{4L2}\\
T_{12,14}T_{10,8} &=& T_{10,14}T_{12,8}\,, \quad \mbox{for (8-10-14-12)}
\,,\label{4L3}\\
T_{16,18}T_{10,8} &=& T_{10,18}T_{16,8}\,, \quad \mbox{for (8-10-18-16)}
\,,\label{4L4}\\
T_{16,18}T_{13,11} &=& T_{13,18}T_{16,11}\,, \quad \mbox{for (11-13-18-16)}
\,,\label{4L5}\\
T_{15,17}T_{14,12} &=& T_{14,17}T_{15,12}\,, \quad \mbox{for (12-14-17-15)}
\,.\label{4L6}
\eeqa
These relations can be used to eliminate all 12 observables of the dash-dot-dot group. 
Furthermore, one finds four 6-point-loops. Two of them, (7-9-13-18-16-11) and 
(8-10-14-17-15-12) contain two dash-dot-dot-lines and do not lead to new relations, 
in fact the corresponding relations 
can be obtained by a proper multiplication of the relations in 
(\ref{4L2}) and (\ref{4L5}), and (\ref{4L3}) and (\ref{4L6}), respectively. Of the other 
two, which involve only solid lines (4) and dashed lines (2), one is a new condition 
\beq
T_{15,12}T_{8,16}T_{11,7} = T_{11,16}T_{8,12}T_{15,7}\,,
\eeq
allowing to eliminate two observables of the dashed-line-group while the second one,
\beq
T_{14,17}T_{9,13}T_{18,10} = T_{18,13}T_{9,17}T_{14,10}\,,
\eeq
is equivalent to the former one, because the corresponding relation is obtained from 
the former one by a proper multiplication of all 4-loop conditions. This then leaves 
the minimal number of 22 observables for describing all interference terms and 
only one further observable is needed to fix the last $t$-matrix element. However, 
one will find then discrete ambiguities due to the elimination of observables using 
the above quadratic relations. We note in passing that the result obtained for 
panel (a) of Fig.\ \ref{connpat1} with regard to the elimination of all observables 
of the dash-dot-dot-line group and two of the dashed-line group is also valid 
for the other five panels in Fig.\ \ref{connpat1} as well as for the corresponding 
connected diagrams constructed with the other solid-line groups of Fig.\ \ref{figThel}.

\newpage

\begin{table}
\caption{Cartesian polarization components of the outgoing nucleons and 
their division into the $A$-type and $B$-type observables.}

\begin{tabular}{ccccccccccccccccc}
$X$ & $1$ & $y_1$ & $y_2$ & $xx$ & $yy$ & $zz$ & $xz$ & $zx$ & 
$x_1$ & $x_2$ & $z_1$ & $z_2$ & $xy$ & $yx$ & $yz$ & $zy$ 
\\[1.ex]
\tableline
&&&&&&&&\\[-2.ex]
type & $A$ & $A$ & $A$ & $A$ & $A$ & $A$ & $A$ & $A$ & 
$B$ & $B$ & $B$ & $B$ & $B$ & $B$ & $B$ & $B$ 
\\
\end{tabular}
\label{tab1}
\end{table}

\begin{table}
\caption{Transformation from cartesian to spherical polarization components 
of the outgoing nucleons.}

\begin{tabular}{ccccc}
 $\alpha$ & $s_\alpha^{00}$ & $s_\alpha^{11}$ & $s_\alpha^{10}$ & 
$s_\alpha^{1-1}$\\
\tableline
 0 & 1 &  0            & 0 & 0 \\
 1 & 0 &  $-1/\sqrt{2}$ & 0 & $1/\sqrt{2}$ \\
 2 & 0 &  $i/\sqrt{2}$  & 0 & $i/\sqrt{2}$ \\
 3 & 0 &  0            & 1 & 0\\
\end{tabular}
\label{tabcartsph}
\end{table}

\begin{table}
\caption{Nomenclature for the diagrammatic representation of groups of 
longitudinal observables in Fig.\ \protect{\ref{figLhel}} 
determining the interference terms of $t$-matrix elements for the helicity basis, 
where $f_L^{IM}(X)$ is represented by $X^{IM}$.}

\begin{tabular}{ccl}
Panel & Line type & Observables 
\\
\tableline
(a) & solid & $x_1^{10}$, $x_2^{22}$, $y_1^{00}$, $y_1^{20}$, $xz^{00}$, $xz^{20}$\\
(b) & solid & $1^{11}$, $1^{21}$, $z_1^{11}$, $z_2^{11}$, $z_1^{21}$, $z_2^{21}$, 
    $zz^{11}$, $zz^{21}$ \\
(c) & solid & $1^{22}$, $z_1^{22}$, $z_2^{22}$, $zz^{22}$ \\
    & dashed & $x_2^{10}$, $x_1^{22}$, $y_1^{22}$, $xz^{22}$ \\
(d) & solid & $x_1^{11}$, $x_2^{11}$, $x_1^{21}$, $x_2^{21}$, 
    $y_1^{11}$, $y_1^{21}$, $xz^{11}$, $xz^{21}$.\\
\end{tabular}
\label{tabgroupLhel}
\end{table}

\begin{table}
\caption{Nomenclature for the diagrammatic representation of groups of 
longitudinal observables in Fig.\ \protect{\ref{figLhyb}} 
determining the interference terms of $t$-matrix elements for the hybrid basis, 
where $f_L^{IM}(X)$ is represented by $X^{IM}$.}

\begin{tabular}{ccl}
Panel & Line type & Observables 
\\
\tableline
(a) & solid & $1^{20}$, $1^{21}$, $1^{22}$, $y_1^{20}$, $y_1^{21}$, $y_1^{22}$\\ 
    & dashed & $xz^{20}$, $xz^{21}$, $xz^{22}$, $zz^{20}$, $zz^{21}$, $zz^{22}$\\ 
(b) & solid & $x_2^{10}$, $x_2^{11}$, $x_2^{21}$, $x_2^{22}$, 
    $z_2^{10}$, $z_2^{11}$, $z_2^{21}$, $z_2^{22}$\\
(c) & solid & $xz^{00}$, $xz^{11}$, $xz^{20}$, $xz^{22}$, 
    $zz^{00}$, $zz^{11}$, $zz^{20}$, $zz^{22}$\\ 
(d) & solid & $x_1^{10}$, $x_1^{11}$, $x_1^{21}$, $x_1^{22}$, 
    $z_1^{10}$, $z_1^{11}$, $z_1^{21}$, $z_1^{22}$\\
\end{tabular}
\label{tabgroupLhyb}
\end{table}

\begin{table}
\caption{Nomenclature for the diagrammatic representation of groups of 
longitudinal observables in Fig.\ \protect{\ref{figLstan}} 
determining the interference terms of $t$-matrix elements for the standard basis, 
where $f_L^{IM}(X)$ is represented by $X^{IM}$. A group in brackets 
with the line type as subscript indicates a subgroup determining the corresponding 
interference term.}

\begin{tabular}{ccl}
Panel & Line type & Observables 
\\
\tableline
(a) & solid & $1^{22}$, $z_1^{22}$, $z_2^{22}$, $zz^{22}$\\ 
    & dashed & $z_1^{10}$, $z_2^{10}$, $z_1^{22}$, $z_2^{22}$\\
(b) & solid & $x_1^{10}$, $x_2^{10}$. $x_1^{22}$, $x_2^{22}$, 
   $(y_1^{00}$, $y_1^{20}$, $xz^{00}$, $xz^{20})_{\mbox{dashed}}$, 
   $y_1^{22}$, $xz^{22}$\\
(c) & solid & $(x_1^{11}$, $x_2^{11}$, $x_1^{21}$, $x_2^{21})_{\mbox{dashed}}$, 
    $(y_1^{11}$, $y_1^{21}$, $xz^{11}$, $xz^{21})_{\mbox{dotted}}$\\ 
(d) & solid & $(1^{11}$, $1^{21}$, $zz^{11}$, $zz^{21})_{\mbox{dashed}}$, 
    $(z_1^{11}$, $z_2^{11}$, $z_1^{21}$, $z_2^{21})_{\mbox{dotted}}$\\
\end{tabular}
\label{tabgroupLstan}
\end{table}

\begin{table}
\caption{Nomenclature for the diagrammatic representation of groups of 
longitudinal observables in Fig.\ \protect{\ref{figLhel4}} determining the 
interference terms of $t$-matrix elements for the transformed helicity basis, 
where $f_L^{IM}(X)$ is represented by $X^{IM}$.}

\begin{tabular}{ccl}
Panel & Line type & Observables 
\\
\tableline
(a) & solid & $xz^{00}$, $xz^{20}$, $xz^{22}$, $zz^{00}$, $zz^{20}$, $zz^{22}$\\ 
(b) & solid & $1^{11}$, $1^{21}$, $y_1^{11}$, $y_1^{21}$\\ 
    & dashed & $xz^{11}$, $xz^{21}$, $zz^{11}$, $zz^{21}$ \\
(c) & solid & $x_1^{10}$, $x_1^{22}$, $z_1^{10}$, $z_1^{22}$ \\
    & dashed & $x_2^{10}$, $x_2^{22}$, $z_2^{10}$, $z_2^{22}$ \\
(d) & solid & $x_1^{11}$, $x_1^{21}$, $z_1^{11}$, $z_1^{21}$ \\
    & dashed & $x_2^{11}$, $x_2^{21}$, $z_2^{11}$, $z_2^{21}$ \\
\end{tabular}
\label{tabgroupLhel4}
\end{table}

\begin{table}
\caption{Numbering of independent longitudinal $t$-matrix elements 
$(\lambda=0)$ of $d(e,e'N)N$ for the helicity basis.}

\begin{tabular}{crrrrrr}
$j$ & $1$ & $2$ & $3$ & $4$ & $5$ & $6$ 
\\
\tableline
&&&&&&\\[-2.ex]
$\lambda_p$ & $\frac{1}{2}$ & $\frac{1}{2}$ & $\frac{1}{2}$ & 
$\frac{1}{2}$ & $\frac{1}{2}$ & $\frac{1}{2}$ 
\\[1.ex] 
$\lambda_n$ & $-\frac{1}{2}$ & $\frac{1}{2}$ & $-\frac{1}{2}$ & 
$\frac{1}{2}$ & $-\frac{1}{2}$ & $\frac{1}{2}$ 
\\[1.ex] 
$\lambda_d$ & $0$ & $0$ & $-1$ & $-1$ & $1$ & $1$ 
\\
\end{tabular}
\label{tabhelL}
\end{table}

\begin{table}
\caption{Numbering of independent transverse $t$-matrix elements 
$(\lambda=1)$ of $d(e,e'N)N$ for the helicity basis.}

\begin{tabular}{crrrrrrrrrrrr}
$j$ & $7$ & $8$ & $9$ & $10$ & $11$ & $12$ & $13$ & $14$ & $15$ & $16$ & $17$ & $18$ 
\\
\tableline
&&&&&&\\[-2.ex]
$\lambda_p$ 
& $\frac{1}{2}$ & $\frac{1}{2}$ & $-\frac{1}{2}$ & $-\frac{1}{2}$ 
& $\frac{1}{2}$ & $\frac{1}{2}$ & $-\frac{1}{2}$ & $-\frac{1}{2}$ 
& $\frac{1}{2}$ & $\frac{1}{2}$ & $-\frac{1}{2}$ & $-\frac{1}{2}$ 
\\[1.ex] 
$\lambda_n$ 
& $\frac{1}{2}$ & $-\frac{1}{2}$ & $\frac{1}{2}$ & $-\frac{1}{2}$ 
& $\frac{1}{2}$ & $-\frac{1}{2}$ & $\frac{1}{2}$ & $-\frac{1}{2}$ 
& $\frac{1}{2}$ & $-\frac{1}{2}$ & $\frac{1}{2}$ & $-\frac{1}{2}$ 
\\[1.ex] 
$\lambda_d$ & $0$ & $0$ & $0$ & $0$ & $1$ & $1$ & $1$ & $1$ 
& $-1$ & $-1$& $-1$ & $-1$ 
\\
\end{tabular}
\label{tabhelT}
\end{table}

\vspace*{1cm}
\begin{table}
\caption{Numbering of independent longitudinal $t$-matrix elements 
$(\lambda=0)$ of $d(e,e'N)N$ for the transformed helicity basis with $R_f=R_y$
and $R_d=R_0=R(0,0,0)$.}

\begin{tabular}{crrrrrr}
$j$ & $1$ & $2$ & $3$ & $4$ & $5$ & $6$ 
\\
\tableline
&&&&&&\\[-2.ex]
$\lambda_p$ & $\frac{1}{2}$ & $-\frac{1}{2}$ & $\frac{1}{2}$ & 
$\frac{1}{2}$ & $-\frac{1}{2}$ & $-\frac{1}{2}$ 
\\[1.ex] 
$\lambda_n$ & $\frac{1}{2}$ & $-\frac{1}{2}$ & $\frac{1}{2}$ & 
$-\frac{1}{2}$ & $\frac{1}{2}$ & $-\frac{1}{2}$ 
\\[1.ex] 
$\lambda_d$ & $0$ & $0$ & $1$ & $1$ & $1$ & $1$ 
\\
\end{tabular}
\label{indLhel4}
\end{table}

\begin{table}
\caption{Exact values for the real and imaginary parts of the 
longitudinal $t$-matrix elements and their 
average values obtained with the experimental simulation for cases A
and B in units of 10 fm. The number in brackets denotes the number of check 
observables. Enumeration of t-matrix elements according to Table 
\protect{\ref{tabhelL}}.}

\begin{tabular}{cccccc}
  & exact  & case A (1) & case A (2) & case B (1) & 
case B (2)\\
\tableline
$\Re e(t_1)$ &  0.1360 &  0.1316 &  0.1339 &  0.1349 &  0.1364 \\ 
$\Re e(t_2)$ & -0.0660 & -0.0681 & -0.0722 & -0.0660 & -0.0668 \\
$\Im m(t_2)$ &  0.0385 &  0.0308 &  0.0389 &  0.0381 &  0.0392 \\
$\Re e(t_3)$ &  0.1285 &  0.1265 &  0.1178 &  0.1260 &  0.1266 \\
$\Im m(t_3)$ & -0.2090 & -0.1918 & -0.2053 & -0.2067 & -0.2088 \\
$\Re e(t_4)$ & -0.1620 & -0.1491 & -0.1631 & -0.1625 & -0.1622 \\
$\Im m(t_4)$ &  0.0310 &  0.0428 &  0.0353 &  0.0314 &  0.0321 \\
$\Re e(t_5)$ & -0.0460 & -0.0469 & -0.0402 & -0.0456 & -0.0451 \\ 
$\Im m(t_5)$ &  0.0338 &  0.0274 &  0.0317 &  0.0344 &  0.0342 \\
$\Re e(t_6)$ &  0.1850 &  0.1842 &  0.1867 &  0.1834 &  0.1852 \\
$\Im m(t_6)$ & -0.0488 & -0.0444 & -0.0560 & -0.0484 & -0.0496 \\ 
\end{tabular}
\label{tab1exp}
\end{table}

\begin{table}
\caption{Standard deviation in the experimental simulation of the 
longitudinal $t$-matrix elements from the average values of Table 
\ref{tab1exp} in units of 10 fm, notation as in Table \protect{\ref{tab1exp}}.}

\begin{tabular}{ccccc}
  & case A (1) & case A (2) & case B (1) & case B (2)\\
\tableline

$\Re e(t_1)$ & 0.0208 & 0.0103 & 0.0119 & 0.0013 \\ 
$\Re e(t_2)$ & 0.0188 & 0.0109 & 0.0066 & 0.0019 \\
$\Im m(t_2)$ & 0.0249 & 0.0068 & 0.0108 & 0.0013 \\
$\Re e(t_3)$ & 0.0344 & 0.0148 & 0.0145 & 0.0043 \\
$\Im m(t_3)$ & 0.0418 & 0.0204 & 0.0242 & 0.0023 \\
$\Re e(t_4)$ & 0.0468 & 0.0259 & 0.0161 & 0.0019 \\
$\Im m(t_4)$ & 0.0356 & 0.0188 & 0.0216 & 0.0033 \\
$\Re e(t_5)$ & 0.0312 & 0.0168 & 0.0138 & 0.0026 \\ 
$\Im m(t_5)$ & 0.0313 & 0.0197 & 0.0073 & 0.0025 \\
$\Re e(t_6)$ & 0.0321 & 0.0154 & 0.0201 & 0.0024 \\
$\Im m(t_6)$ & 0.0308 & 0.0199 & 0.0166 & 0.0031 \\ 
\end{tabular}
\label{tab2exp}
\end{table}

\begin{table}
\caption{Selected complete sets for the numerical solution of the transverse
$t$-matrix elements.}

\begin{tabular}{cccccccccccc}
$1^{00}$ & $1^{11}$ & $1^{20}$ & $1^{21}$ & $1^{22}$ & $1^{\prime\,10}$ &
  $1^{\prime\,11}$ & $1^{\prime\,21}$ & $1^{\prime\,22}$ & $y_1^{00}$ &
  $y_1^{11}$ & $y_1^{20}$ \\  
& $y_1^{21}$ & $y_1^{22}$ & $y_1^{\prime\,10}$ & $y_1^{\prime\,11}$ &
  $y_1^{\prime\,21}$ & $y_1^{\prime\,22}$ & $x_1^{10}$ & $x_1^{11}$ &
  $x_1^{\prime\,00}$ & $x_1^{\prime\,11}$ & $x_1^{\prime\,20}$ \\
\tableline
$1^{00}$ & $1^{11}$ & $1^{21}$ & $1^{\prime\,11}$ & $1^{\prime\,21}$ &
 $z_1^{11}$ & $z_1^{21}$ & $z_1^{\prime\,11}$ & $z_1^{\prime\,21}$ &
 $x_2^{22}$ & $x_2^{\prime\,22}$ & $y_2^{22}$ \\
& $y_2^{\prime\,22}$ & $z_2^{11}$ & $z_2^{21}$ & $z_2^{\prime\,11}$ &
  $z_2^{\prime\,21}$ & $zx^{22}$ & $zy^{22}$ & $zz^{11}$ & $zz^{21}$ &
  $zz^{\prime\,11}$ & $zz^{\prime\,21}$ \\
\end{tabular}
\label{tabexpT}
\end{table}

\begin{table}
\caption{Listing of bounds for longitudinal and transverse structure 
functions $f_L^{IM}(X)$ and $f_T^{IM}(X)$.}
\begin{tabular}{cccc}
$I$ & $M$ & $\lambda_{min}/2$ & $\lambda_{max}/2$\\
\tableline
\multicolumn{4}{c}{$X=1$} \\
\tableline
1 & 1 & $-\sqrt{3}$ & $\sqrt{3}$ \\ 
2 & 0 & $-\sqrt{2}$ & $1/\sqrt{2}$ \\ 
2 & 1 & $-\sqrt{3}$ & $\sqrt{3}$ \\ 
2 & 2 & $-\sqrt{3}$ & $\sqrt{3}$ \\ 
\tableline
\multicolumn{4}{c}{$X=y_1,\,y_2,\,xx,\,xz,\,zx$} \\
\tableline
0 & 0 & $-1$ & $1$ \\ 
1 & 1 & $-\sqrt{3}$ & $\sqrt{3}$ \\ 
2 & 0 & $-\sqrt{2}$ & $\sqrt{2}$ \\ 
2 & 1 & $-\sqrt{3}$ & $\sqrt{3}$ \\ 
2 & 2 & $-\sqrt{3}$ & $\sqrt{3}$ \\ 
\tableline
\multicolumn{4}{c}{$X=x_1,\,x_2,\,z_1,\,z_2$} \\
\tableline
1 & 0 & $-\sqrt{3/2}$ & $\sqrt{3/2}$ \\ 
1 & 1 & $-\sqrt{3}$ & $\sqrt{3}$ \\ 
2 & 1 & $-\sqrt{3}$ & $\sqrt{3}$ \\ 
2 & 2 & $-\sqrt{3}$ & $\sqrt{3}$ \\ 
\end{tabular}
\label{tableL-T}
\end{table}

\begin{table}
\caption{Listing of bounds for transverse and transverse interference 
structure functions $f_T^{\prime\,IM}(X)$ and $f_{TT}^{IM\pm}(X)$.}
\begin{tabular}{cccc}
$I$ & $M$ & $\lambda_{min}/2$ & $\lambda_{max}/2$\\
\tableline
\multicolumn{4}{c}{$f_T^{\prime\,IM}(X),\,f_{TT}^{IM-}(X):\,
X=1,\,y_1,\,y_2,\,xx,\,xz,\,zx;$} \\
\multicolumn{4}{c}{$f_{TT}^{IM+}(X):\,X=x_1,\,x_2,\,z_1,\,z_2$} \\
\tableline
1 & 0 & $-\sqrt{3/2}$ & $\sqrt{3/2}$ \\ 
1 & 1 & $-\sqrt{3}$ & $\sqrt{3}$ \\ 
2 & 1 & $-\sqrt{3}$ & $\sqrt{3}$ \\ 
2 & 2 & $-\sqrt{3}$ & $\sqrt{3}$ \\ 
\tableline
\multicolumn{4}{c}{$f_T^{\prime\,IM}(X),\,f_{TT}^{IM-}(X):\,
X=x_1,\,x_2,\,z_1,\,z_2;$} \\
\multicolumn{4}{c}{$f_{TT}^{IM+}(X):\,
X=1,\,y_1,\,y_2,\,xx,\,xz,\,zx$} \\
\tableline
0 & 0 & $-1$ & $1$ \\ 
1 & 1 & $-\sqrt{3}$ & $\sqrt{3}$ \\ 
2 & 0 & $-\sqrt{2}$ & $\sqrt{2}$ \\ 
2 & 1 & $-\sqrt{3}$ & $\sqrt{3}$ \\ 
2 & 2 & $-\sqrt{3}$ & $\sqrt{3}$ \\ 
\end{tabular}
\label{tableTp-TT}
\end{table}

\begin{table}
\caption{Listing of bounds for longitudinal-transverse interference 
structure functions $f_{LT}^{IM\pm}(X)$ and $f_{LT}^{\prime\,IM\pm}(X)$.}
\begin{tabular}{cccc}
$I$ & $M$ & $\lambda_{min}/2$ & $\lambda_{max}/2$\\
\tableline
\multicolumn{4}{c}{$f_{LT}^{IM+}(X),\,f_{LT}^{\prime\,IM+}(X):\,
X=1,\,y_1,\,y_2,\,xx,\,xz,\,zx;$} \\
\multicolumn{4}{c}{$f_{LT}^{IM-}(X),\,f_{LT}^{\prime\,IM-}(X):\,
X=x_1,\,x_2,\,z_1,\,z_2$} \\
\tableline
0 & 0 & $-\sqrt{2}$ & $\sqrt{2}$ \\ 
1 & 1 & $-\sqrt{6}$ & $\sqrt{6}$ \\ 
2 & 0 & $-2$ & $2$ \\ 
2 & 1 & $-\sqrt{6}$ & $\sqrt{6}$ \\ 
2 & 2 & $-\sqrt{6}$ & $\sqrt{6}$ \\ 
\tableline
\multicolumn{4}{c}{$f_{LT}^{IM+}(X),\,f_{LT}^{\prime\,IM+}(X):\,
X=x_1,\,x_2,\,z_1,\,z_2;$} \\
\multicolumn{4}{c}{$f_{LT}^{IM-}(X),\,f_{LT}^{\prime\,IM-}(X):\,
X=1,\,y_1,\,y_2,\,xx,\,xz,\,zx$} \\
\tableline
1 & 0 & $-\sqrt{3}$ & $\sqrt{3}$ \\ 
1 & 1 & $-\sqrt{6}$ & $\sqrt{6}$ \\ 
2 & 1 & $-\sqrt{6}$ & $\sqrt{6}$ \\ 
2 & 2 & $-\sqrt{6}$ & $\sqrt{6}$ \\ 
\end{tabular}
\label{tableLT-LTp}
\end{table}

\begin{table}
\caption{Nomenclature for the diagrammatic representation of groups of 
transverse observables in Fig.\ \protect{\ref{figThel}} 
determining the interference terms of 
$t$-matrix elements for the helicity basis, where $f_T^{(\prime)\,IM}(X)$ 
is represented by $X^{(\prime)\,IM}$.}

\begin{tabular}{ccl}
Panel & Line type & Observables 
\\
\tableline
(a) & solid & $1^{11}$, $1^{21}$, $z_1^{11}$, $z_1^{21}$, 
           $z_2^{11}$, $z_2^{21}$, $zz^{11}$, $zz^{21}$, 
\\
&&
           $1^{\prime\,11}$, $1^{\prime\,21}$, $z_1^{\prime\,11}$, 
           $z_1^{\prime\,21}$, $z_2^{\prime\,11}$, $z_2^{\prime\,21}$, 
           $zz^{\prime\,11}$, $zz^{\prime\,21}$
\\
 & dashed & $1^{22}$, $z_1^{22}$, $z_2^{22}$, $zz^{22}$, 
            $1^{\prime\,22}$, $z_1^{\prime\,22}$, $z_2^{\prime\,22}$, 
            $zz^{\prime\,22}$
\\
\tableline
 (b) & solid & $x_2^{11}$, $x_2^{21}$, $y_2^{11}$, $y_2^{21}$, $zx^{11}$, 
               $zx^{21}$, $zy^{11}$, $zy^{21}$,
           
\\
&&
           $x_2^{\prime\,11}$, $x_2^{\prime\,21}$, $y_2^{\prime\,11}$,  
           $y_2^{\prime\,21}$, $zx^{\prime\,11}$, $zx^{\prime\,21}$, 
           $zy^{\prime\,11}$, $zy^{\prime\,21}$
\\
 & dashed & $x_2^{22}$, $y_2^{22}$, $zx^{22}$, $zy^{22}$, $x_2^{\prime\,22}$, 
            $y_2^{\prime\,22}$, $zx^{\prime\,22}$, $zy^{\prime\,22}$
\\
 & dash-dot-dot & $x_2^{10}$, $y_2^{00}$, $y_2^{20}$, $zx^{00}$, 
                  $zx^{20}$, $zy^{10}$, 
                  \\
&&
                  $x_2^{\prime\,00}$, $x_2^{\prime\,20}$, $y_2^{\prime\,10}$, 
                  $zx^{\prime\,10}$, $zy^{\prime\,00}$, $zy^{\prime\,20}$
\\
\tableline
 (c) & solid & $xx^{11}$, $xx^{21}$, $xy^{11}$, $xy^{21}$, $yx^{11}$, $yx^{21}$, 
               $yy^{11}$, $yy^{21}$, 
\\
&&
           $xx^{\prime\,11}$, $xx^{\prime\,21}$, $xy^{\prime\,11}$, $xy^{\prime\,21}$, 
           $yx^{\prime\,11}$, $yx^{\prime\,21}$, $yy^{\prime\,11}$, $yy^{\prime\,21}$
\\
 & dashed & $xx^{22}$, $xy^{22}$, $yx^{22}$, $yy^{22}$, $xx^{\prime\,22}$, 
           $xy^{\prime\,22}$, $yx^{\prime\,22}$, $yy^{\prime\,22}$
\\
 & dash-dot-dot & $xx^{00}$, $xx^{20}$, $xy^{10}$, $yx^{10}$, $yy^{00}$, $yy^{20}$, 
\\
&&
                  $yy^{\prime\,10}$, $xx^{\prime\,10}$, $xy^{\prime\,00}$, 
                  $xy^{\prime\,20}$, $yx^{\prime\,00}$, $yx^{\prime\,20}$
\\
\tableline
 (d) & solid & $x_1^{11}$, $x_1^{21}$, $y_1^{11}$, $y_1^{21}$, $xz^{11}$, 
               $xz^{21}$, $yz^{11}$, $yz^{21}$, 
           
\\
&&
           $x_1^{\prime\,11}$, $x_1^{\prime\,21}$, $y_1^{\prime\,11}$, 
           $y_1^{\prime\,21}$, $xz^{\prime\,11}$, $xz^{\prime\,21}$, 
           $yz^{\prime\,11}$, $yz^{\prime\,21}$ 
\\
 & dashed & $x_1^{22}$, $y_1^{22}$, $xz^{22}$, $yz^{22}$, $x_1^{\prime\,22}$, 
            $y_1^{\prime\,22}$, $xz^{\prime\,22}$, $yz^{\prime\,22}$ 
\\
 & dash-dot-dot & $x_1^{10}$, $y_1^{00}$, $y_1^{20}$, $xz^{00}$, $xz^{20}$, 
                  $yz^{10}$, \\
&&
                  $x_1^{\prime\,00}$, $x_1^{\prime\,20}$, $y_1^{\prime\,10}$, 
                  $xz^{\prime\,10}$, $yz^{\prime\,00}$, $yz^{\prime\,20}$ 
                  
\\
\end{tabular}
\label{tabgroupThel}
\end{table}

\begin{table}
\caption{Nomenclature for the diagrammatic representation of groups of 
transverse observables in Fig.\ \protect{\ref{figThel}} 
determining the interference terms of 
$t$-matrix elements for the transformed helicity basis, where $f_T^{(\prime)\,IM}(X)$ 
is represented by $X^{(\prime)\,IM}$.}

\begin{tabular}{ccl}
Panel & Line type & Observables 
\\
\tableline
(a) & solid & $1^{11}$, $1^{21}$, $y_1^{11}$, $y_1^{21}$, 
           $y_2^{11}$, $y_2^{21}$, $yy^{11}$, $yy^{21}$, 
\\
&&
           $1^{\prime\,11}$, $1^{\prime\,21}$, $y_1^{\prime\,11}$, $y_1^{\prime\,21}$, 
           $y_2^{\prime\,11}$, $y_2^{\prime\,21}$, $yy^{\prime\,11}$, $yy^{\prime\,21}$
\\
 & dashed & $1^{22}$, $y_1^{22}$, $y_2^{22}$,  $yy^{22}$, $1^{\prime\,22}$, 
            $y_1^{\prime\,22}$, $y_2^{\prime\,22}$, $yy^{\prime\,22}$ 
\\
\tableline
 (b) & solid & $x_2^{11}$, $x_2^{21}$, $z_2^{11}$, $z_2^{21}$, $yx^{11}$, 
               $yx^{21}$, $yz^{11}$, $yz^{21}$, 
\\
&&
           $x_2^{\prime\,11}$, $x_2^{\prime\,21}$, $z_2^{\prime\,11}$,  
           $z_2^{\prime\,21}$, $yx^{\prime\,11}$, $yx^{\prime\,21}$, 
           $yz^{\prime\,11}$, $yz^{\prime\,21}$
\\
 & dashed & $x_2^{22}$, $z_2^{22}$, $yx^{22}$, $yz^{22}$, $x_2^{\prime\,22}$, 
            $z_2^{\prime\,22}$, $yx^{\prime\,22}$, $yz^{\prime\,22}$ 
\\
 & dash-dot-dot & $x_2^{10}$, $z_2^{10}$, $yx^{10}$, $yz^{10}$, 
                  $x_2^{\prime\,00}$, $x_2^{\prime\,20}$, 
                  $z_2^{\prime\,00}$, $z_2^{\prime\,20}$, \\
&&
                  $yx^{\prime\,00}$, $yx^{\prime\,20}$, 
                  $yz^{\prime\,00}$, $yz^{\prime\,20}$
                  
\\
\tableline
 (c) & solid & $xx^{11}$, $xx^{21}$, $xz^{11}$, $xz^{21}$, 
           $zx^{11}$, $zx^{21}$, $zz^{11}$, $zz^{21}$, 
\\
&&
           $xx^{\prime\,11}$, $xx^{\prime\,21}$, $xz^{\prime\,11}$, $xz^{\prime\,21}$, 
           $zx^{\prime\,11}$, $zx^{\prime\,21}$, $zz^{\prime\,11}$, $zz^{\prime\,21}$
\\
 & dashed & $xx^{22}$, $xz^{22}$, $zx^{22}$, $zz^{22}$,  
            $xx^{\prime\,22}$, $xz^{\prime\,22}$, $zx^{\prime\,22}$, 
            $zz^{\prime\,22}$
\\
 & dash-dot-dot & $xx^{00}$, $xx^{20}$, $xz^{00}$, $xz^{20}$, $zx^{00}$, $zx^{20}$,       
                  $zz^{00}$, $zz^{20}$, \\
&&
                  $xx^{\prime\,10}$, $xz^{\prime\,10}$, 
                  $zx^{\prime\,10}$, $zz^{\prime\,10}$,
\\
\tableline
 (d) & solid & $x_1^{11}$, $x_1^{21}$, $z_1^{11}$, $z_1^{21}$, $xy^{11}$, 
               $xy^{21}$, $zy^{11}$, $zy^{21}$, 
\\
&&
           $x_1^{\prime\,11}$, $x_1^{\prime\,21}$, $z_1^{\prime\,11}$, 
           $z_1^{\prime\,21}$, $xy^{\prime\,11}$, $xy^{\prime\,21}$, 
           $zy^{\prime\,11}$, $zy^{\prime\,21}$, 
\\
 & dashed & $x_1^{22}$, $z_1^{22}$, $xy^{22}$, $zy^{22}$,  
            $x_1^{\prime\,22}$, $z_1^{\prime\,22}$$xy^{\prime\,22}$, 
            $zy^{\prime\,22}$
\\
 & dash-dot-dot & $x_1^{10}$, $z_1^{10}$, $xy^{10}$, $zy^{10}$, 
                  $x_1^{\prime\,00}$, $x_1^{\prime\,20}$,
                  $z_1^{\prime\,00}$, $z_1^{\prime\,20}$,\\
&&
                  $xy^{\prime\,00}$, $xy^{\prime\,20}$, 
                  $zy^{\prime\,00}$, $zy^{\prime\,20}$
\\
\end{tabular}
\label{tabgroupThel4}
\end{table}

\end {document}